\documentclass[12pt,letterpaper]{JHEP3}
\usepackage{amsmath,amssymb,amsfonts,xspace}
\usepackage{epsfig}
\usepackage{epstopdf}

\numberwithin{equation}{section}

\def\Im{\,{\rm Im}\,}
\def\Re{\,{\rm Re}\,}

\def\({\left(}
\def\){\right)}
\def\[{\left[}
\def\]{\right]}

\def\hf{\frac{1}{2}}

\newcommand{\de}{\mathrm{d}}

\newcommand{\I}{\mathrm{i}}
\newcommand{\e}{\mathrm{e}}

\newcommand{\p}{\partial}

\newcommand{\cB}{\mathcal{B}}
\newcommand{\cF}{\mathcal{F}}

\newcommand{\cS}{\mathcal{S}}

\newcommand{\cK}{\mathcal{K}}

\newcommand{\cM}{\mathcal{M}}
\newcommand{\cW}{\mathcal{W}}
\newcommand{\cN}{\mathcal{N}}
\newcommand{\cE}{\mathcal{E}}
\newcommand{\cX}{\mathcal{X}}

\newcommand{\cJ}{\mathcal{J}}

\DeclareSymbolFont{AMSa}{U}{msa}{m}{n}
\DeclareSymbolFont{AMSb}{U}{msb}{m}{n}
\DeclareMathSymbol{\fieldR}{\mathalpha}{AMSb}{"52}

\newcommand{\N}{{\mathcal N}}
\newcommand{\kahler}{{K\"ahler}\xspace}
\newcommand{\hk}{{hyperk\"ahler}\xspace}
\newcommand{\qk}{{quaternion-K\"ahler}\xspace}

\newcommand{\cZ}{\mathcal{Z}}

\newcommand{\cO}{\mathcal{O}}
\newcommand{\cH}{\mathcal{H}}
\newcommand{\cU}{\mathcal{U}}

\newcommand{\pa}{\partial}
\newcommand{\nn}{\nonumber}

\newcommand{\eps}{\epsilon}
\newcommand{\IR}{\mathbb{R}}
\newcommand{\IC}{\mathbb{C}}
\newcommand{\IZ}{\mathbb{Z}}
\newcommand{\IH}{\mathbb{H}}

\newcommand{\Tr}{\mbox{Tr}}

\newcommand{\tzeta}{{\tilde\zeta}}

\newcommand{\txi}{{\tilde\xi}}
\newcommand{\tnu}{{\tilde\nu}}
\newcommand{\tchi}{{\tilde\chi}}

\newcommand{\CP}{\mathbb{P}}

\newcommand{\beq}{\begin{eqnarray}}
\newcommand{\eeq}{\end{eqnarray}}
\def\be{\begin{equation}}
\def\ee{\end{equation}}
\def\ba{\begin{align}}
\def\ea{\end{align}}
\def\bse{\begin{subequations}}
\def\ese{\end{subequations}}
\newcommand{\bea}[2]{\be\label{#2}\begin{array}{#1}}
\newcommand{\eea}{\end{array}\ee}

\def\ba{\bar a}

\def\bs{\bar s}
\def\bz{\bar z}

\def\bZ{\bar Z}

\def\bX{\bar X}
\def\bF{\bar F}

\def\talp{\tilde\alpha}

\def\Sij#1{S^{[#1]}}

\def\Hij#1{H^{[#1]}}

\def\zp{z_{+}}
\def\zm{z_{-}}
\def\zpm{z_{\pm}}

\def\XXint#1#2#3{{\setbox0=\hbox{$#1{#2#3}{\int}$}
\vcenter{\hbox{$#2#3$}}\kern-.5\wd0}}

\def\hH{\hat H}
\def\cij#1{c}
\def\ci#1{c}

\def\ba{\bar a}

\def\bt{\bar t}

\def\bz{\bar z}

\def\tc{\tilde c}

\def\varpi{t}

\def\CY{\mathfrak{Y}}
\def\CYm{\hat\CY}

\def\CY{\mathfrak{Y}}
\def\CYm{\mathfrak{\hat Y}}
\def\Kf{\mathfrak{S}}

\def\Ma{M}
\def\Mi{M}
\def\Mn{M}

\def\hh{\hat h}

\title{Heterotic-type II duality in twistor space}

\preprint{L2C:12-162 \\ CERN-PH-TH/2012-259\\arXiv:1210.3037v2}

\author{
Sergei Alexandrov$^{1,2}$, Boris Pioline$^{3,4}$
\\
$^1$ {\it Universit\'e Montpellier 2, Laboratoire Charles Coulomb UMR 5221, F-34095,
Montpellier, France}\\

$^2$ {\it CNRS, Laboratoire Charles Coulomb UMR 5221, F-34095,
Montpellier, France}\\

$^3$ {\it CERN PH-TH,
Case C01600, CERN, CH-1211 Geneva 23, Switzerland}\\

$^4$ {\it Laboratoire de Physique Th\'eorique et Hautes
Energies, CNRS UMR 7589, \\
Universit\'e Pierre et Marie Curie,
4 place Jussieu, 75252 Paris cedex 05, France} \\

\vspace*{2mm} {\tt e-mail:
\email{salexand@univ-montp2.fr},
\email{boris.pioline@cern.ch}
}

\vspace*{-3mm}

}

\abstract{Heterotic string theory compactified on a $K3$ surface times $T^2$ is believed to be
equivalent to type II string theory on a suitable Calabi-Yau threefold.
In particular, it must share the same hypermultiplet moduli space.
Building on the known twistorial description on
the type II side, and on recent progress on the map between type II and
heterotic moduli in the limit where both the type II and
heterotic strings become classical, we provide a
new twistorial construction of the hypermultiplet moduli space
in this limit which is adapted to the symmetries of the heterotic string.
We also take steps towards understanding  the twistorial description for
heterotic worldsheet instanton corrections
away from the classical limit.
As a spin-off, we obtain a twistorial description of a class of automorphic
forms of $SO(4,n,\IZ)$ obtained by Borcherds' lift.}

\begin{document}

\section{Introduction}

String vacua with $\cN=2$ supersymmetry in 4 dimensions can be constructed in many different ways,
in particular by compactifying the type IIA (type IIB, respectively) string theory on
a Calabi-Yau threefold $\CY$ ($\CYm$, respectively), or the $E_8\times E_8$ heterotic string
on a K3 surface $\Kf$ times a torus $T^2$.
For suitable choices of the manifolds $\CY,\CYm,\Kf$,
and of the gauge bundle on the heterotic side,
these constructions are believed to be equivalent, and related by non-perturbative
dualities \cite{Witten:1995ex,Kachru:1995wm,Ferrara:1995yx}.
Strong evidence for this claim comes
from the matching of various protected couplings in the low energy effective action,
and a detailed comparison of the spectrum of BPS states.

So far, most of these tests have focused on quantities which depend only on
vector multiplet (VM) scalars. Since the type IIB dilaton and the \kahler moduli of $\CYm$
are hypermultiplet scalars (HM), such quantities can typically be computed exactly at
zero coupling, infinite volume on the type IIB side, and then compared
with the expected result on the heterotic side. The latter depends
on the heterotic dilaton and Narain moduli of the torus $T^2$, and typically receives
a finite set of computable perturbative corrections, together with an infinite
series of five-brane instanton corrections predicted by duality (see e.g.
\cite{Aspinwall:2000fd} and references therein).

The implications of string dualities for the hypermultiplet sector of $\cN=2$
string vacua have been far less studied, with some notable exceptions
\cite{Aspinwall:1998bw,Aspinwall:1999xs,Aspinwall:2000fd,Halmagyi:2007wi}.
A technical reason is that hypermultiplet scalars are valued in a \qk (QK) space \cite{Bagger:1983tt},
a class of geometries rather more sophisticated than the special \kahler
geometry of the vector multiplet scalars. Fortunately, powerful
twistorial techniques have been developed
in the mathematics and physics literature
\cite{MR664330,MR1327157,MR1096180,Lindstrom:1983rt,Karlhede:1984vr,Hitchin:1986ea,Lindstrom:1987ks,
Ivanov:1995cy,deWit:2001dj,Alexandrov:2008ds,Alexandrov:2008nk,Lindstrom:2008gs},
providing an efficient computable framework
to deal with such spaces.

More fundamentally, none of the dual descriptions above has so far provided a simple
framework to compute quantum corrections to the HM moduli space.
On the type II side, the HM moduli space $\cM_{\rm H}$ receives
D-brane and NS5-brane instanton corrections \cite{Becker:1995kb}.
Although much progress has been achieved recently in describing these instanton corrections
\cite{RoblesLlana:2006is,Alexandrov:2006hx,Saueressig:2007dr,RoblesLlana:2007ae,
Alexandrov:2008gh,Pioline:2009qt,Bao:2009fg,Alexandrov:2009zh,Alexandrov:2009qq,Alexandrov:2010np,
Alexandrov:2010ca,Alexandrov:2011ac,Alexandrov:2012bu,Alexandrov:2012au}
(see \cite{Alexandrov:2011va} for a review), a complete understanding is still lacking.
In particular, the computation of the generalized Donaldson-Thomas (DT) invariants which govern D-instanton
contributions appears to be out of reach in general.
On the heterotic side, since the dilaton is a part of the VM moduli, the same HM moduli
space $\cM_{\rm H}$ should be entirely computable at zero string coupling, within the worldsheet
$(0,4)$ superconformal sigma model on $\Kf$. However, the effect of worldsheet instantons
\cite{Witten:1999eg} is ill-understood except near certain singularities
\cite{Witten:1999fq,Rozali:1999va}. Moreover, the HM moduli space
on the heterotic side includes, beyond the usual geometric and B-field moduli of $\Kf$,
data about the $E_8\times E_8$ gauge bundle on $\Kf$ \cite{Friedman:1997yq},
which is considerably more intricate than the usual Narain moduli space of flat bundles on $T^2$.

Our aim in this work is to carry the twistorial techniques which were successfully
used in the study of the HM moduli space in type II Calabi-Yau vacua over to the heterotic side.
For simplicity, we focus on perturbative heterotic compactifications
on an elliptic K3 surface $\cE\to \Kf \to \cB$, conjecturally dual to type IIB compactifications on
a K3-fibered CY threefold $\Sigma \to \CYm \to \CP$ \cite{Aspinwall:1995vk}.
We further assume that all bundle moduli are fixed,\footnote{In the first release of this work,
we allowed for the existence of bundle moduli, and focused on the base of the fibration
$\cM_{\rm HK}\to \cM_{\rm H}\to \cM_{\rm red}$, where $\cM_{\rm HK}$ is the gauge bundle
moduli space for fixed geometric and B-field moduli of $\Kf$, parametrized by $\cM_{\rm red}$.
We no longer believe that this fibration holds true, but rather that  $\cM_{\rm HK}$ is fibered
over the space of geometric moduli of $\Kf$, while the B-field moduli live in a torus bundle
fibered over the total space of this fibration. A detailed study of this structure will appear
in \cite{Louis:2013}.}
as happens for example for pointlike
instantons at singularities of $\Kf$. On the type IIB side, the absence of such
bundle moduli requires that the K3-fibration has no reducible singular fibers \cite{Aspinwall:1998bw}.
In the limit where  the area of the fiber $\cE$ and base $\cB$
are both scaled to infinity on the heterotic side, it is known that $\cM_{\rm H}$
reduces to the orthogonal Wolf space \cite{MR0185554}
\be
\label{Wolf}
\cW(n) = \frac{SO(4,n)}{SO(4)\times SO(n)}\, ,
\ee
where $n$ is the number of unobstructed deformations of
$\Kf$. The duality map between heterotic and type II variables in this limit,
first discussed in  \cite{Aspinwall:1998bw} and completed recently in \cite{Louis:2011aa},
is recalled in \eqref{mirmaptree} below.

In this work, we develop a twistorial description of the HM moduli space $\cM_{\rm H}$
which is adapted to the symmetries of the heterotic string on $\Kf$, in particular
to the automorphism group $SO(3,n-1,\IZ)$ of the lattice of (unobstructed)
two-cycles on $\Kf$ and to large gauge transformations of the $B$-field.
Mathematically, this heterotic twistorial description
is based on the fact that the \hk cone (HKC) of the orthogonal Wolf space
$\cW(n)$ is the \hk quotient of $\mathbb{H}^{n+4}$ under $SU(2)$ at
zero-level \cite{MR872143}.
Rather than starting from this somewhat abstract statement, we begin with the
now standard  twistorial description of the HM moduli space
on the type IIB side \cite{Neitzke:2007ke, Alexandrov:2008nk,Alexandrov:2008gh} and determine the moment maps
of the isometries $\pa_{B_I}$ associated to Peccei-Quinn symmetries along the $B$-field moduli
(as well as the `opposite' isometries $_{B_I}\pa$, related to the former by the Cartan involution).
We show that those provide (generalized) Darboux coordinates
on the heterotic side,
and take a simple covariant form \eqref{result-etamu} in terms of the heterotic fields.
The relations \eqref{momapA}, \eqref{momapB} between these heterotic Darboux coordinates
and the standard type II coordinates \eqref{gentwi} may be considered as a lift of the duality map of \cite{Louis:2011aa}
to twistor space.


We further investigate the  twistorial description of heterotic worldsheet
instanton effects which correct the metric on $\cM_{\rm H}$ away from the classical limit.
Although we cannot give a full answer to this question, we do find a simple twistorial description
for the simplified but closely related problem of finding zero-modes of the Baston operator,
a second-order differential operator acting on the space of functions on any QK manifold.
The function \eqref{H1Om2} generating the zero modes via Penrose transform is one of our main results.
As a spin-off, we find that a class of automorphic forms under $SO(4,n,\IZ)$ obtained
by Borcherds' lift  is  annihilated by the Baston operator, and therefore has
a simple twistorial description generated by the same holomorphic function.

The remainder of this note is organized as follows. In \S\ref{sec_rev} we recall
the basic structure of the HM moduli space on the type IIB and heterotic side, and
the duality map which relates them in the classical limit. In \S\ref{sec-classZ},
after briefly reviewing the twistorial construction of the HM moduli space on the type II
side, we provide an alternative construction which is adapted to the symmetries of the
heterotic string, and explain its relation to the standard type II construction. In \S\ref{sec-wsinst},
we start investigating the twistorial description of heterotic worldsheet instanton corrections,
in a linear approximation around the classical HM metric. We close in \S\ref{sec_disc}
with a summary and directions for further research. Some technical computations
are relegated to Appendices \ref{ap-quotient} and \ref{ap-instact},
while Appendix \ref{sec_theta} discusses the application to automorphic forms obtained by Borcherds' lift.

\section{HM moduli space in Heterotic/type II dual pairs \label{sec_rev}}

In this section we review the basic structure of the hypermultiplet moduli space in type IIB
string theory compactified on a K3-fibered CY threefold $\CYm$ and in the dual
$E_8\times E_8$ heterotic string compactified on an elliptically fibered K3 surface $\Kf$,
as well as the duality map between them in the classical limit. We follow \cite{Louis:2011aa}
except for minor changes in conventions.

\subsection{Moduli space in type IIB on a K3-fibered CY threefold \label{sec_modiib}}

At zeroth order in the string coupling constant, the HM moduli space $\cM_{\rm H}$
in type IIB string theory compactified on a threefold $\CYm$ is obtained from
the \kahler moduli space $\cM_{\rm K}$ of  $\CYm$ by the $c$-map construction
\cite{Cecotti:1989qn,Ferrara:1989ik}. $\cM_{\rm K}$
itself is described by a holomorphic prepotential $F(Z^\Lambda)$ homogeneous of degree
2 in the projective \kahler moduli $Z^\Lambda$ ($\Lambda=0,\dots, h_{1,1}(\CYm)$).
The homogeneous ratios  $z^a=Z^a/Z^0$ ($a=1,\dots, h_{1,1}(\CYm))$ measure the periods of the
complexified \kahler form $B+\I \cJ$ on a  basis $\gamma^a$
of the homology lattice $H_2(\CYm,\IZ)$, dual to a basis $\gamma_a$ of $H_4(\CYm,\IZ)$
under the cup product. As usual we denote by $\gamma^0$ and $\gamma_0$
the generators of $H_0(\CYm)$ and $H_6(\CYm)$, and equip $H_{\rm even}(\CYm,\IZ)$
with an integer symplectic pairing such that
$\gamma^\Lambda, \gamma_\Lambda$ with $\Lambda=0,a$ form a symplectic basis.

In order for heterotic-type II duality to apply, it is necessary to assume that $\CYm$ is
a K3 fibration $\Sigma\rightarrow \CYm \rightarrow \CP$ by algebraic K3 surfaces
$\Sigma$ \cite{Aspinwall:1995vk}.
As a result, the  lattice $H_4(\CYm,\IZ)$ has a distinguished element $\gamma_s=\Sigma$ given by the
generic K3 fiber. We assume that the fibration has a global section, such that $\gamma_s$
is dual to the base $\gamma^s=\CP\in H_2(\CYm,\IZ)$ of the fibration. The remaining generators of
$H_4(\CYm,\IZ)$ consist of the `Picard divisors' $\gamma_i$ ($i=1,\dots, n-2$, the offset
by $2$ is for later convenience)
obtained by sweeping an algebraic cycle  in ${\rm Pic}(\Sigma)$ along a monodromy invariant
cycle on the base, and of `exceptional divisors' $\gamma_\alpha$ ($\alpha=1,\dots,  h_{1,1}(\CYm)-n+2$)
associated to reducible singular fibers.
Here ${\rm Pic}(\Sigma)=H_2(\Sigma,\IZ)\cap H_{1,1}(\Sigma,\IR)$ is the Picard lattice,
i.e. the sublattice of algebraic 2-cycles inside $H_2(\Sigma,\IZ)$, and $n-2$ is its rank.
For simplicity, we assume that no reducible bad fibers are present, so $n=h_{1,1}(\CYm)+2$.
We denote by $\eta_{ij}$ the signature $(1,n-3)$
intersection form on ${\rm Pic}(\Sigma)$, and by $\kappa_{abc}=(\gamma_a,\gamma_b,\gamma_c)$
the intersection product on $H_4(\CYm,\IZ)$. A key fact is that $\kappa_{sbc}$ vanishes unless $b,c$ correspond to
Picard divisors $\gamma_i, \gamma_j$, in which case it equals $\eta_{ij}$.
In the limit where the base is very large, the prepotential thus takes the form
\be
\label{FclasG}
F(Z^\Lambda)= -\frac{Z^s\eta_{ij}Z^iZ^j}{2Z^0}
+f(Z^0,Z^i) + \cO(e^{2\pi\I Z^s/Z^0}) \, ,
\ee
where $f(Z^0,Z^i)$ is independent of $Z^s$ and includes classical contributions from the
intersection product of Picard divisors, the usual 4-loop $\alpha'$
correction proportional to the Euler number  of $\CYm$, and worldsheet instantons wrapping
Picard  divisors. The last term in \eqref{FclasG} corresponds to
worldsheet instantons wrapping the base $\CP$,
which are exponentially suppressed when its area $\Im z^s$ is very large.

Starting from the prepotential \eqref{FclasG}, the $c$-map
construction \cite{Ferrara:1989ik} produces the QK metric
\be
\begin{split}
\de s_{\rm H}^2=& \, \frac{1}{r^2}\,\de r^2
-\frac{1 }{2r}\, (\Im\cN)^{\Lambda\Sigma}\(\de\tzeta_\Lambda -\cN_{\Lambda\Lambda'}\de\zeta^{\Lambda'}\)
\(\de\tzeta_\Sigma -\bar\cN_{\Sigma\Sigma'}\de\zeta^{\Sigma'}\)
\\ &
+ \frac{1}{16r^2}\(\de\sigma+\tzeta_\Lambda\de\zeta^\Lambda-\zeta^\Lambda\de\tzeta_\Lambda\)^2
+4\cK_{a\bar b}\de z^a \de \bz^{\bar b}  ,
\end{split}
\label{hypmettree}
\ee
where $\cK$ is the K\"ahler potential on the special K\"ahler moduli space $\cM_{\rm K}$,
\be
\cK=-\log K\, ,
\qquad
K=-2\Im\(\bZ^\Lambda F_\Lambda\)\, ,
\qquad
\cK_{a\bar b} = \pa_{z^a}\pa_{{\bar z}^{\bar b}} \cK
\label{Kahlerpot_spec}
\ee
and $\cN_{\Lambda\Sigma}$ is the `Weil period matrix'
\be
\cN_{\Lambda\Sigma} =\bF_{\Lambda\Sigma} +
2\I\, \frac{ \Im F_{\Lambda\Lambda'}Z^{\Lambda'} \Im F_{\Sigma\Sigma'}Z^{\Sigma'}}
{Z^\Xi \Im F_{\Xi\Xi'}Z^{\Xi'}}\,  .
\ee
As usual, we denote $F_\Lambda=\pa_{Z^\Lambda}F,\ F_{\Lambda\Sigma}=\pa_{Z^\Sigma}F_\Lambda$.
The coordinate $r$ is related to the  type IIB string coupling in 4  dimensions by
$r=1/g_s^2$,
while $(\zeta^\Lambda,\tzeta_\Lambda)$ correspond to the periods of the RR gauge fields
on $H_{\rm even}(\CYm,\IZ)$, and $\sigma$ is the NS axion dual to the Kalb-Ramond two-form
$B$ in four dimensions (see \cite{Alexandrov:2008gh} for more details).

In the limit where the area $\Im z^s$ of the base is very large, the prepotential \eqref{FclasG}
reduces to the leading, cubic term, while the \kahler potential \eqref{Kahlerpot_spec}  takes the form
\be
\cK=-\log(\Re s)-\log\(-\eta_{ij}(z^i-\bz^i)(z^j-\bz^j)\)-\log |Z^0|^2,
\ee
where we have defined $z^s=\I s$.
In order to recognize $\cM_{\rm H}$ as a symmetric space,
and to prepare for the comparison with the heterotic side, it is convenient
to perform a symplectic transformation
\be
\label{symp-tr}
S\ :\ (Z^s,F_s)\ \mapsto\ (F_s,-Z^s)\, .
\ee
Denoting by
\be
(X^A,G_A) = S\cdot (Z^\Lambda,F_\Lambda),
\qquad
(c^A,\tc_A)=S\cdot (\zeta^\Lambda,\tzeta_\Lambda)
\ee
the new symplectic vectors after the transformation $S$,
we have
\be
X^A=\Bigl(X^0,-\frac{X^0}{2}\,\eta_{ij}z^i z^j,X^0 z^i\Bigr),
\qquad
G_A=-\I s\, \eta_{AB} X^B,
\label{newbase}
\ee
where we introduced the following  signature $(2,n-2)$ intersection product on the
lattice ${\rm Pic}_Q(\Sigma) \equiv \IZ\, \gamma^0 \oplus \IZ \gamma^s \oplus {\rm Pic}(\Sigma)$,
\be
\label{etaABij}
\eta_{AB}=\( \begin{array}{ccc}
\, 0 \ & 1\,  & 0
\\
\,1\  & 0\, & 0
\\
\,0\ & 0\, & \eta_{ij}
\end{array}\) .
\ee
We refer to the rank $n$ lattice ${\rm Pic}_Q(\Sigma)$ as the `quantum Picard lattice'.

It is crucial to note that the fields $X^A$ are not independent, but satisfy
the quadratic constraint $\eta_{AB}X^A X^B=0$. In particular, the symplectic vector $(X^A,G_A)$
does not follow from a prepotential.
The \kahler potential and period matrix in this basis are given by  \cite{Ceresole:1995jg}
\be
\begin{split}
K= &\, (s+\bs)\eta_{AB}X^A\bX^B,
\\
\cN_{AB}
=&\,
-\frac{\I}{2}\((s-\bs)\eta_{AB}+(s+\bs)M_{AB}\)
=\I \bs\eta_{AB}-\I(s+\bs)X_{AB},
\end{split}
\label{newN}
\ee
where we introduced
\be
X_{AB}=\frac{X_A\bX_B+\bX_A X_B}{\eta^{CD}X_C\bX_D},
\qquad
\Ma_{AB}=-\eta_{AB}+2X_{AB}.
\label{matXM}
\ee
Note that $(\Ma^{-1})^{AB}=\Ma^{AB}$ where the indices are raised and lowered by the matrix $\eta_{AB}$.
Thus the matrix $M_{AB}$ parametrizes the symmetric space $\frac{SO(2,n-2)}{SO(2)\times SO(n-2)}$. Noting that
\be
\de \Ma_{AB} \de \Ma^{AB}=-16\, \cK_{i\bar\jmath}\, \de t^i\de\bt^j\, ,
\label{diffM}
\ee
we find that the metric on $\cM_{\rm H}$ can be rewritten as
\be
\begin{split}
\de s^2=& \frac{\de r^2}{r^2}-\frac{1}{4}(\de \Ma)^2+\frac{4\, \de s \, \de\bar s}{(s+\bs)^2}
+\ \frac{\Ma^{AB}}{r(s+\bs)}\(\de\tc_A -\cN_{AA'}\de c^{A'}\)
\(\de\tc_B -\bar\cN_{BB'}\de c^{B'}\)
\\ &
+\frac{1}{16 r^2}\(\de\sigma +\tc_A\de c^A-c^A\de \tc_A\)^2,
\end{split}
\label{hypmet2}
\ee
This identifies the metric \eqref{hypmettree} in the  limit $\Re(s)\to +\infty$
as the $c$-map of the special \kahler space
$\frac{SL(2)}{U(1)} \times \frac{SO(2,n-2)}{SO(2)\times SO(n-2)}$, where the first factor
is parametrized by the \kahler modulus $s$ of the base $\CP$ . It is well known, and will become apparent
shortly, that the
$c$-map of this space is the orthogonal Wolf space
$\cW(n)$ \cite{deWit:1992wf}.

\subsection{Moduli space in heterotic string theory on K3}

Let us now consider $E_8\times E_8$ heterotic  string theory compactified on a K3 surface $\Kf$ times $T^2$.
We restrict to perturbative vacua without NS5-branes, and assume that the gauge bundle is flat on $T^2$.
The VM multiplet moduli space is parametrized by the \kahler and complex moduli $T,U$ of the torus,
the holonomies $y^m$  of the Cartan generators of the unbroken part $E_8\times E_8$,
together with the complexified four-dimensional string coupling $\psi + \I/g_h^2$
(where $\psi$ denotes the pseudoscalar dual to the heterotic Kalb-Ramond two-form $B$).
As is well known, it receives tree-level, one-loop and instanton corrections from heterotic
five branes wrapping $K3\times T^2$.

In contrast, the HM moduli space is independent
of the heterotic string coupling, and is therefore entirely determined by
the (0,4) superconformal field theory at tree-level \cite{Ferrara:1995yx}.
It parametrizes the \hk metric, $B$-field and $E_8\times E_8$ gauge bundle on $\Kf$.
In particular, it is independent of the radius of the torus $T^2$, and unaffected
under decompactification to 6 dimensions.
In order to preserve $\cN=2$ supersymmetries, the gauge bundle on $\Kf$ must have anti-self
dual curvature $F=-\star F,\ F'=-\star F'$ (where $F,F'$ correspond to the two $E_8$ factors).
Due to the Bianchi identity for the Kalb-Ramond field strength
\be
\de H = \alpha' \left[ \Tr(R\wedge R)  - \Tr(F\wedge F) - \Tr(F'\wedge F') \right] ,
\ee
the second Chern classes of the two $E_8$ bundles add up to $c_2+c_2'=\chi(\Kf)=24$.
In general, some of the deformations of the HK metric
may be obstructed by the requirement that $(F,F')$ stays anti-self dual.
For now we assume that this is not the case, and return to this point below.

As is well known from the study of $(4,4)$ superconformal sigma models \cite{Seiberg:1988pf},
the moduli space of HK metrics on a K3 surface $\Kf$ is
\be
\label{MEinstein}
\IR^+_{\rho} \times \[\frac{SO(3,19)}{SO(3)\times SO(19)}\]_{\gamma_I^x}\, ,
\ee
where the first factor corresponds to the overall volume $V_{\Kf}=e^{-\rho}$ in
heterotic string units, and the second factor is parametrized by the periods of
the triplet of \kahler forms  $\cJ^x$, $x=1,2,3$ along a basis
$\tau_I$ ($I=1,\dots ,22$) of  $H_2(\Kf,\IZ)$, Poincar\'e dual to a basis
$\omega^I$  of $H^2(\Kf,\IZ)$:
\be
\label{defgIx}
\gamma_I^x = e^{\rho/2} \, \eta_{IJ}\, \int_{\Kf} \cJ^x \, \wedge \omega^J\, ,
\qquad
\cJ^x = e^{-\rho/2} \gamma_I^x \omega^I\, ,
\ee
where the second equation holds up to exact forms.
Here we have lowered the index using the (inverse of the) signature $(3,19)$
intersection form $\eta^{IJ}$ on $H^2(\Kf,\IZ)$, defined by
\be
\int_{\tau_I}\omega^J=\delta_I^J,
\qquad
\int_{\Kf} \omega^I\,\wedge \omega^J = \eta^{IJ} .
\ee
It is important to note that the periods  $\gamma_I^x$ are subject to the orthogonality condition
\be
\label{Jxyrho}
\cJ^x\wedge \cJ^y=2\delta^{xy}\, {\rm Vol}_{\Kf}
\quad
\Leftrightarrow
\quad
\eta^{IJ} \gamma_I^x \gamma_J^y = 2\delta^{xy}\, .
\ee
Moreover a $SU(2)$ rotation of $\cJ^x$ leaves the HK metric on $\Kf$ unaffected. The metric on
\eqref{MEinstein} is the $SO(3,19)$ invariant metric
\be
\label{dsEinstein}
\de s^2_{\rm HK} = \frac{1}{2}\,\de\rho^2 - \frac14 \de \Mi_{IJ} \de \Mi^{IJ} ,
\ee
where $\Mi_{IJ}$ is the symmetric coset representative
\be
\Mi_{IJ}=-\eta_{IJ}+\gamma^x_I\gamma^x_J,
\qquad
\Mi^{IJ} = \eta^{IK} \eta^{JL} \Mi_{KL} .
\ee

On top of these metric moduli, the SCFT on $\Kf$ is also parametrized by the periods of
the $B$ field
\be
B_I = \eta_{IJ}\, \int_{\Kf} B\, \wedge \omega^J\, ,
\qquad
B = B_I\, \omega^I\, .
\ee
Due to large gauge transformations, the periods $B_I$ are defined up to integer shifts
and live in a flat torus bundle of rank 22 over the moduli space \eqref{MEinstein}
of HK  metrics on $\Kf$.
In the large volume limit, the total metric is invariant under translations along the
torus fibers, and takes the form
\be
\label{dsSCFT}
\de s^2_{\rm SCFT} = \frac{1}{2}\,\de\rho^2 - \frac14\, \de \Mi_{IJ} \de \Mi^{IJ} +
e^\rho \, \Mi^{IJ} \de B_I\,\de B_J \, .
\ee
To recognize this metric, we introduce the signature $(4,20)$ quadratic form
\be
\eta_{MN}=\( \begin{array}{ccc}
0 & 0 & -1 \\
0 & \eta_{IJ} & 0 \\
-1 & 0 & 0
\end{array}\) ,
\label{defetaMN}
\ee
where the indices $M,N$ run over $\flat,I,\sharp$,
and the matrix
\be
\Mn_{MN} =
\begin{pmatrix}
1 & & \\
B_J & \delta^I_J & \\
\tfrac12 B_I \eta^{IJ} B_J & B^I & 1
\end{pmatrix}
\cdot
\begin{pmatrix}
e^\rho & & \\
& \Mi_{IJ} & \\
& & e^{-\rho}
\end{pmatrix}\cdot
\begin{pmatrix}
1 &  B_I & \tfrac12 B_I \eta^{IJ} B_J\\
& \delta_I^J & B^J\\
& & 1
\end{pmatrix}
\label{MatrixM}
\ee
belonging to the $SO(4,20)$ symmetry group.
The tree-level metric then takes the manifestly invariant form
\be
\de s^2_{\rm SCFT} = -\frac{1}{4}\, \de \Mn_{MN} \, \de \Mn^{MN}\, ,
\ee
which is recognized as the invariant QK metric on the orthogonal Wolf space $\cW(n)$ \eqref{Wolf} with $n=20$.

This is not, however, the HM moduli space of heterotic strings on $\Kf$, as one in general needs to
incorporate the gauge bundle moduli. As mentioned above, some of the deformations
of the HK metric on $\Kf$ may be obstructed by the anti-self-duality condition on $(F,F')$.
This occurs e.g. in the case of small instantons stuck at ALE singularities of $\Kf$,
which have no bundle moduli, obstruct the blow up modes of the metric and freeze the
periods $\gamma_I^x$ and $B$-field $B_I$ around
the corresponding vanishing cycles to zero \cite{Witten:1995gx,Aspinwall:1997ye}.
We shall
assume that, by this or some other mechanism, all bundle moduli are fixed, so that
the HM moduli space in the large volume limit is given by
\be
\cM_{\rm H}=\cW(n) \ ,\quad n\leq 20\ ,
\ee
parametrizing the QK moduli space of (possibly singular) HK metrics and
$B$ fields on $\Kf$ unobstructed by the (possibly singular) gauge bundle.


\subsection{Heterotic string on elliptic K3 surface}

In order to relate the heterotic and type II pictures, it is useful to restrict to the case where
the K3 surface $\Kf$ is algebraic and fibered by elliptic curves, $\cE\rightarrow \Kf\rightarrow \cB$.
We again assume that the fibration has a global section, which we denote by $\cB$, so that the Picard lattice
${\rm Pic}(\Kf)=H_2(\Kf,\IZ)\cap H_{1,1}(\Kf,\IR)$
decomposes into $\Gamma_{1,1}\oplus {\rm Pic}'(\Kf)$, where $\Gamma_{1,1}$ is generated by
the base $\cB$ and the generic fiber $\cE$, with signature $(1,1)$ intersection
form ${\scriptsize \begin{pmatrix} 2 & 1 \\ 1 & 0 \end{pmatrix}}$. The two-cycles
in ${\rm Pic}'(\Kf)$ are associated to ADE singularities of $\Kf$, and the corresponding
periods $\gamma_I^x$ are frozen due to the gauge bundle. The metric and $B$-field moduli
then arise from the remaining  2-cycles in
\be
\label{TransQ}
{\rm Trans}_Q(\Kf)\equiv \IZ \, \cB \oplus \IZ\, \cE \oplus {\rm Trans}(\Kf)\subset H_2(\Kf,\IZ) ,
\ee
where $ {\rm Trans}(\Kf)$ is the orthogonal complement of  $ {\rm Pic}(\Kf)$ inside $H_2(\Kf,\IZ)$,
known as the transcendental lattice.
We refer to  ${\rm Trans}_Q(\Kf)$
as the `quantum transcendental lattice'. In the basis $\tau_I$ ($I=1,2,A$) such that
$\tau_1=\cB-\cE,\ \tau_2=\cE$ and  $\tau_A$ is a basis of ${\rm Trans}(\Kf)$,
the intersection form on \eqref{TransQ} is given by
\be
\eta_{IJ}=\( \begin{array}{ccc}
\, 0\ & 1\, & 0
\\
\,1\ & 0\, & 0
\\
\,0\ & 0\, & \eta_{AB}
\end{array}\) ,
\ee
where $\eta_{AB}$ is the signature $(2,n-2)$ intersection form on ${\rm Trans}(\Kf)$.
In order to match the moduli space on the type II side, it is clear that one should have
\be
{\rm Trans}(\Kf) = {\rm Pic}_Q(\Sigma) ,
\ee
where ${\rm Pic}_Q(\Sigma)$ is the quantum Picard lattice defined above \eqref{etaABij}.
In particular the Picard numbers of $\Kf$
and $\Sigma$ must add up\footnote{In fact, $\Kf$ and $\Sigma$ are related by mirror symmetry,
which exchanges the quantum Picard lattice with the transcendental lattice
(see e.g. \cite{Aspinwall:1996mn} and references therein).}
to 20.

Choosing $\cJ_3$ as the \kahler form for the complex structure in which $\Kf$ is algebraic,
such that $\int_{\cE} \cJ =0$, where $\cJ\equiv \tfrac12(\cJ^1+\I \cJ^2)$,
the complex structure on $\Kf$ can be parametrized by the complex periods
\be
\int_{\tau_A} \cJ = \frac{e^{-\rho/2}}{\sqrt{X^A \eta_{AB} \bar X^B}}\,  X_A \, ,
\ee
subject to the quadratic condition
\be
\eta^{AB} \, \int_{\tau_A} \cJ \, \int_{\tau_B} \cJ \propto X_A \eta^{AB} X_B= 0\, ,
\ee
which follows from \eqref{Jxyrho}.
The complex variables $X_A$, modulo real rescalings $X_A\mapsto \lambda X_A$
and subject to the quadratic constraint above, parametrize the Grassmannian $\frac{SO(2,n-2)}{SO(2)\times SO(n-2)}$,
which is indeed the moduli space of complex structures on algebraic K3 surfaces with Picard number $22-n$
\cite{Aspinwall:1996mn}.

The \kahler moduli on the other hand are
measured by the periods of $\cJ_3$, which we parametrize by real variables $v_A$ and $R$
according to
\be
\begin{split}
\int_{\cE} \cJ^3 =& \int_{\Kf} \cJ^3 \wedge  \omega^1 =  e^{-(\rho-R)/2} ,
\\
\int_{\tau_A} \cJ^3 = &\eta_{AB} \int_{\Kf} \cJ^3 \wedge \omega^B =e^{-(\rho-R)/2}\, v_A .
\end{split}
\ee
The periods of $\cJ$ and $\cJ_3$ on the remaining cycle $\cB$ are then determined by
\eqref{Jxyrho}:
\be
\begin{split}
\int_{\cB} \cJ =&  -\frac{e^{-\rho/2}}{\sqrt{X^A \eta_{AB} \bar X^B}}\, v_A X^A\, ,
\\
\int_{\cB} \cJ^3 =& \int_{\Kf} \cJ^3 \wedge ( \omega^2-\omega^1 )
= e^{-(\rho+R)/2} -  \( \frac{v^2}{2} +1\)  e^{-(\rho-R)/2}\, ,
\end{split}
\ee
where $v^2=v_A \eta^{AB} v_B$.
In terms of the periods $\gamma_I^x$ defined in \eqref{defgIx}, these identifications amount to
taking a $SU(2)$ frame such that
 \be
\begin{split}
&
\gamma_1=-\eta^{AB}\gamma_A v_B,
\qquad\qquad\
\gamma_2=0,
\qquad\quad\
\gamma_A=\frac{X_A}{\sqrt{\eta^{BC}X_B\bX_C}},
\\
&
\gamma_1^3=e^{-R/2} -\frac{ v^2}{2}\, e^{R/2},
\qquad
\gamma_2^3=e^{R/2},
\qquad
\gamma_A^3=e^{R/2} v_A .
\end{split}
\label{param-zetatree}
\ee
In terms of these variables, the invariant metric \eqref{dsEinstein}
on the moduli space of HK metrics may be rewritten as
\be
\label{dsEinstein2}
\de s^2_{\rm HK} = \frac{1}{2}\,\de\rho^2 - \frac14 \biggl(
\de \Ma_{AB} \de \Ma^{AB} -2\de R^2-4 e^R \Ma^{AB}\de v_A \de v_B \biggr),
\ee
where $\Ma_{AB}$ is defined in terms of $X^A$ as in \eqref{matXM}.
Finally, the periods $B_I$ of the Kalb-Ramond two-form on $H_2(\Kf,\IZ)$ decompose into
\be
\label{Bmap}
B_1 =  \int_{\cB+\cE} B \, ,
\qquad
B_2 = \int_{\cE} B \, ,
\qquad
B_A = \int_{\tau_A} B \, .
\ee

\subsection{Classical limit and duality map\label{sec_dualmap}}

After these preparations, we are finally ready to identify the heterotic and type IIB variables.
Comparing the metrics \eqref{dsSCFT} and \eqref{hypmet2} on the HM moduli
space, we find that the two agree
provided \cite{Louis:2011aa}\footnote{To translate to the notations in \cite{Louis:2011aa}, set
$$\( e^R,e^\rho, s,v^A, B_1, B_2 , B_A, \gamma_A^x,\gamma_{1}^x,\gamma_{2}^x\)_{\rm here} =
\(8 e^R,2 e^\rho , 2s,\frac{1}{2\sqrt2} v^A, \frac14 b, 2 b_*, \frac{1}{\sqrt2} B_A,\zeta_A^x,
\frac{1}{2\sqrt2}\zeta_{1}^x,2\sqrt2\zeta_{2}^x\)_{\cite{Louis:2011aa}}
$$}
\be
\begin{split}
&
r=\hf\,e^{-\frac12(\rho+R)} ,
\qquad
s=\e^{-\hf(\rho-R)} -\I B_2 ,
\\
c^A= v^A,&\,
\qquad
\tc_A=B_A-B_2 v_A,
\qquad
\sigma=- 2B_1 - v^A\, B_A .
\end{split}
\label{mirmaptree}
\ee
This identification holds in the regime  where the volume $\Re s=e^{-\frac12(\rho-R)}$
of the $\CP$ base on the type IIB side is large (such that only the first term in \eqref{FclasG}
remains), and the type IIB string coupling $g_s = 1/\sqrt{r} \propto e^{\frac14(\rho+R)}$
is small, such that the type IIB string can be treated classically.
This is achieved by sending $\rho\to-\infty$, keeping other variables fixed.
On the heterotic side, this corresponds to the  `double scaling limit'
where the area of the elliptic fiber $e^{-\frac12(\rho-R)}$ and of the base $e^{-\frac12(\rho+R)}$ are both
scaled to infinity \cite{Friedman:1997yq,Aspinwall:1997ye,Aspinwall:1998bw}.

In this limit, the metric on $\cM_{\rm H}$ admits an isometric action of $SO(4,n)$.
In particular, it is invariant under  continuous shifts of the B-field moduli $B_I$, corresponding to
the grade +1 elements in the grading
\be
\label{deco420}
so(4,n) =  [n+2]_{-1} \oplus ( so(3,n-1) \oplus so(1,1) )_{0} \oplus [n+2]_{+1}
\ee
induced by the decomposition \eqref{defetaMN}. On the type IIB
side, the corresponding Killing vectors become
\be
\label{KilB}
\pa_{B_1} = -2\pa_{\sigma},
\qquad
\pa_{B_2}= -\pa_{\Im s}-c_A \pa_{\tc_A},
\qquad
\pa_{B_A}\ = \pa_{\tc_A} - c^A \pa_{\sigma}.
\ee
Away from the classical limit, these continuous symmetries are in general broken
to discrete translations by heterotic worldsheet instantons \cite{Witten:1999fq}.
On the type IIB side,
continuous symmetries along $B_2$ are broken by worldsheet instantons wrapping
the base $\CP$ of the K3 fibration, while continuous symmetries along
$B_0, B_i, B_s$ are broken to discrete translations by D5-brane instantons wrapping $\CYm$,
D3-brane instantons wrapping Picard divisors $\gamma_i$,
and D1-brane instantons wrapping the base $\CP$, respectively \cite{Alexandrov:2008gh}.
Finally, continuous translations along $B_1$ are broken by NS5-branes wrapping
$\CYm$ on the type IIB side \cite{Alexandrov:2010ca}.
The duality map \eqref{mirmaptree} ensures that
the natural periodicities of the type II variables $(\zeta^A, \tzeta_A,\sigma/2)$
are mapped to integer periodicities of the $B_I$'s and the $v^A$'s.

\section{The duality map in twistor space}
\label{sec-classZ}

Much of the recent progress in understanding quantum corrections to HM moduli spaces
has been achieved by formulating them in twistor space, where the constraints from
QK geometry are easily implemented.
The twistorial formulation used in \cite{Alexandrov:2008gh} was tailored for the
application to type II strings compactified on a CY threefold, and based on the $c$-map structure
of the HM moduli space in the weak coupling limit. Our goal in this section is to provide
an alternative construction of the same HM moduli space which is instead tailored for
the heterotic string compactified on K3, and relate it to the type IIB construction
of \cite{Alexandrov:2008gh}.

\subsection{Brief review of twistorial techniques for QK manifolds}
\label{subsec-twistor}

The main advantage of the twistor approach to \qk geometry is that it provides
an explicit description of any \qk space $\cM$ in terms of a set of holomorphic
functions \cite{Alexandrov:2008nk}. Those may, roughly speaking, be considered
as analogues of the prepotential in special K\"ahler geometry.
If $d$ denotes the quaternionic dimension of $\cM$,
these functions live on a $2d+1$-complex dimensional
bundle $S^2\to \cZ\to \cM$  known as the twistor space.
Despite the fact that $\cM$ in general possesses no complex structure,
$\cZ$ is equipped with a canonical complex contact structure and a real involution which
acts trivially on $\cM$ and as the antipodal map on the $S^2$
fiber \cite{MR664330,MR1327157}. Conversely, provided the fibration
is sufficiently generic (see the references above for the technical details),
any complex contact manifold $\cZ$ equipped with
such a real involution gives rise to a QK space $\cM$.

The central object in the twistor construction is the complex contact structure.
It is constructed from the $SU(2)_R$ part of
the spin connection\footnote{Recall that a QK manifold has by definition restricted
holonomy $Sp(d)\times SU(2)\subset SO(4d)$. We denote by $SU(2)_R$ the second factor
in this decomposition.}
  on $\cM$ as the kernel of the one-form
\be
\label{XDt}
\cX = 8\I\,e^{\Phi}\,\frac{Dt}{t}  ,
\qquad
Dt= \de t+ p_+ -\I p_3 t + p_-  t^2  ,
\ee
where $t$ denotes a stereographic coordinate on $S^2$.
The one-form $Dt$ is of Dolbeault type $(1,0)$ on $\cZ$,
and valued in the $\cO(2)$ line bundle over $S^2$.
The one-form $\cX$ is defined only up to complex rescalings.
We choose the factor $e^{\Phi}$, known as the `contact potential',
such that $\cX$ is holomorphic (i.e. $\bar\pa$-closed). Locally (but not globally), one can always
choose `contact Darboux coordinates' $\talp, \xi^\Lambda, \txi_\Lambda$
on some open patch $\cU$ in $\cZ$ such that
\be
\label{cXdarb}
\cX =\de\talp+ \txi_\Lambda\, \de \xi^\Lambda - \xi^\Lambda \, \de \txi_\Lambda .
\ee
The global complex contact structure, and hence the QK metric on $\cM$,
can then be specified by a set of complex contact transformations between local Darboux
coordinate systems on an open covering $\cU_i$ of $\cZ$. The generating functions
of these transformations, $\Sij{ij}$, are the holomorphic functions mentioned above.
Of course, they must satisfy the obvious consistency conditions
on triple overlaps $\cU_i\cap \cU_j\cap \cU_k$, are defined up to local complex
contact transformations $\Sij{i}$ on each patch, and must also be consistent
with the real involution.

Linear deformations of the QK metric on $\cM$ are described by
linear deformations of the contact transformations $\Sij{ij}$,
therefore by real classes in $H^1(\cZ,\cO(2))$. Likewise, quaternionic isometries of the QK metric on $\cM$
lift to infinitesimal contact transformations $\Sij{i}$, therefore to global sections of $H^0(\cZ,\cO(2))$.
To any such isometry, the moment map construction associates
a holomorphic function $\cO$ on $\cZ$ which generates the infinitesimal contact transformation via
the `contact Poisson action' on holomorphic functions $\cF$ \cite{Alexandrov:2008gh}
\be
\begin{split}
\left[\cO ,\cF\right\}=&\,
-\(\p_{\txi_\Lambda}+\xi^\Lambda\p_{\talp}\) \cO \,
\pa_{\xi^\Lambda} \cF + \(\p_{\xi^\Lambda}-\txi_\Lambda\p_{\talp}\) \cO \,\pa_{\txi_\Lambda} \cF
\\
&\,
+ \(\xi^\Lambda\p_{\xi^\Lambda}+\txi_\Lambda\p_{\txi_\Lambda}-2\) \cO\, \pa_{\talp} \cF .
\end{split}
\label{contaction}
\ee
The function generating the commutator of two Killing vectors
can be obtained by the `contact Poisson bracket' \cite{Alexandrov:2008gh}
\be
\begin{split}
\{ \cO_1,\cO_2 \} = &  -2\cO_1 \pa_{\talp} \cO_2 - \pa_{\talp} \cO_1 (\xi^\Lambda \pa_{\xi^\Lambda}
+\txi_\Lambda \pa_{\txi_\Lambda})
\cO_2 +  \pa_{\xi^\Lambda} \cO_1 \pa_{\txi_\Lambda} \cO_2 - (1\leftrightarrow 2) .
\end{split}
\label{contpoisson}
\ee

For some purposes, in particular for making isometries of $\cM$ manifest,
it is useful to consider the Swann bundle $\cS$, a $\IC^\times$-bundle over $\cZ$
(or $\IC^2/\IZ_2$-bundle over $\cM$) endowed with a \hk cone metric \cite{MR1096180,deWit:2001dj}.
As such it admits a homogeneous complex symplectic form $\omega_{\cS}$,
which can be cast locally in Darboux form
\be
\label{defomS}
\omega_{\cS} = \de \nu^I \wedge \de \tnu_I\, ,
\ee
where $\nu^I$ are taken to transform with degree 1 under the $\IC^\times$ action,
whereas $\tnu_I$ stay inert \cite{Alexandrov:2008nk}.
The QK metric on $\cM$ is now encoded into a set of complex symplectomorphisms between different
local Darboux coordinate patches, which commute with the $\IC^\times$ action.
These symplectomorphisms are simply related to the contact transformations above  by projectivization:
singling out one of the indices $I$, say $1$, and letting $\Lambda$ run over the remaining indices,
one may defining projective coordinates $\nu=\nu^1$, $\xi^\Lambda=\nu^\Lambda/\nu^1$,
$\txi_\Lambda=\tnu_\Lambda$, $\talp=-2\tnu_1-\xi^\Lambda\txi_\Lambda$ such that $\xi^\Lambda,\txi_\Lambda,\talp$
provide local complex contact Darboux coordinates on $\cZ=\cS/\IC^\times$,
while $\nu$ parametrizes the fiber.
The complex symplectic form \eqref{defomS} then decomposes as
\be
\omega_{\cS} = -\hf\(\de \nu\, \wedge\, \cX +\nu\, \de \cX\),
\label{gensymform}
\ee
where $\cX$ is the contact one-form \eqref{cXdarb}.
One may check that the contact Poisson action \eqref{contaction} and bracket \eqref{contpoisson}
are given by $\{ \nu \cO_1, \cO_2\}_{\cS}$ and
$\{\nu \cO_1,\nu \cO_2\}_{\cS}$, where $\{ \cdot, \cdot \}_{\cS}$ is the Poisson bracket on $\cS$
for the symplectic form \eqref{defomS}.
One advantage of the Swann bundle is that any quaternionic isometry of
$\cM$ lifts to a triholomorphic action on $\cS$.

\subsection{Type II Darboux coordinates}
\label{subsec-typeIIZ}

We now briefly recall how the QK  metric on the type IIB HM moduli space \eqref{hypmettree}
can be cast in this twistorial framework. The construction works for any $c$-map metric,
irrespective of the form of the prepotential.
One introduces Darboux coordinates \cite{Neitzke:2007ke,Alexandrov:2008nk}
\be
\label{gentwi}
\begin{array}{rcl}
\xi^\Lambda &=& \zeta^\Lambda + \frac12\,\tau_2
\left( \varpi^{-1} Z^{\Lambda} -\varpi \,\bZ^{\Lambda}  \right) ,
\\
\txi_\Lambda &=& \tzeta_\Lambda + \frac12\,\tau_2
\left( \varpi^{-1} F_\Lambda-\varpi \,\bF_\Lambda \right) ,
\\
\talp&=& \sigma + \frac12\,\tau_2
\left(\varpi^{-1} W -\varpi \,\bar W \right),
\end{array}
\ee
where, in the gauge $Z^0=1$,
\be
\label{defW}
\tau_2 = 4\sqrt{r}\, e^{\cK/2} ,
\qquad
W = F_\Lambda(z) \zeta^\Lambda - Z^\Lambda \tzeta_\Lambda .
\ee
Fixing the coordinates $r,Z^\Lambda,\zeta^\Lambda,\tzeta_\Lambda,\sigma$ on
the QK space and varying the fiber coordinate $t\in S^2$, these expressions parametrize
the twistor lines, i.e. the fibers of the projection $\cZ\to \cM$.
Substituting the Darboux coordinates in the contact one-form
\eqref{cXdarb} and choosing $e^\Phi=r$,
one may check that the QK metric reconstructed from
the spin connection $p_\pm, p_3$ in \eqref{XDt} reproduces \eqref{hypmettree}.
The Darboux coordinates above are valid away from the north and south poles $t=0$ and $t=\infty$.
The singularity at those points may be removed by a complex contact transformation generated
by the prepotential $F(\xi^\Lambda)$ \cite{Alexandrov:2008nk}.
It is worth noting that the Darboux coordinates $(\xi^\Lambda,\txi_\Lambda)$ defined in \eqref{gentwi}
transform as a vector under symplectic transformations, while $\talp$ is inert.

Performing the symplectic
rotation \eqref{symp-tr}, which is necessary for later comparison with the heterotic formulation,
the coordinates $\xi^A,\txi_A$ are replaced by
\be
\label{gentwiA}
\begin{array}{rcl}
\chi^A &=& c^A + \frac12\,\tau_2
\left( \varpi^{-1} X^{A} -\varpi \,\bar X^{A}  \right) ,
\\
\tchi_A &=& \tc_A + \frac12\,\tau_2
\left( \varpi^{-1} G_A -\varpi \,\bar G_A \right) ,
\end{array}
\ee
while $\tau_2$, $W=G_A c^A - X^A \tc_A$ and $\talp$ are unaffected. The contact
form after the transformation takes the same form as in \eqref{cXdarb}
\be
\label{cXdarbA}
\cX =\de\talp+ \tchi_A\, \de \chi^A - \chi^A \, \de \tchi_A\,  .
\ee

\subsection{Heterotic moment maps}
\label{subsec-coordetamu}

In order to relate the heterotic and type II variables in \S\ref{sec_dualmap}, a key step was
to identify the Killing vectors  \eqref{KilB} for the commuting isometries corresponding to
shifts of the heterotic $B$-field. Likewise, in order to
understand the structure of the twistor space in heterotic variables, it is important to consider
the moment maps $\eta^I$ associated to the isometries $\pa_{B_I}$. A simple computation
shows that these are given by
\be
\label{momapA}
\eta^1 =1,
\qquad
\eta^2= -\hf\, \eta_{AB} \chi^A \chi^B ,
\qquad
\eta^A=\chi^A .
\ee
A central role will also be played by the moment maps $\mu_I$ associated to the `opposite'
isometries corresponding to the grade $-1$ generators in \eqref{deco420}, which we denote by $_{B_I}\pa$.
While the action of these Killing vectors on the type II variables is unilluminating, their
moment maps can be computed using the known formulae for the isometries of
$c$-map spaces with cubic prepotential \cite{deWit:1990na,deWit:1992wf}, obtaining
\be
\label{momapB}
\begin{split}
\mu_1 =&\, \frac14\,\left( \talp^2
+ \eta_{AB}\chi^A\chi^B\eta^{CD}\tchi_C\tchi_D -(\chi^A\tchi_A)^2 \right) ,
\qquad
\mu_2 = -\hf\,\eta^{AB}\tchi_A \tchi_B ,
\\
\mu_A =& -\hf\( (\talp-\chi^B\tchi_B)\tchi_A+\eta_{AB}\chi^B\eta^{CD}\tchi_C\tchi_D\) .
\end{split}
\ee

A useful check on these expressions is that they are indeed global $\cO(2)$ sections,  i.e.
are quadratic in $t$ (modulo an overall factor of $1/t$) when expressed in terms of the coordinates
on $\cM_{\rm H}\times S^2$, and satisfy the algebra
\be
\{\eta_I, \eta_J\} =  \{\mu_I, \mu_J\}=0\, ,
\qquad
\{ \eta_I, \mu_J \} = -h_{IJ} - h \eta_{IJ}
\ee
under the contact Poisson bracket \eqref{contpoisson}. These results are in agreement with the fact that
the isometries $\pa_{B_I}$ (respectively, $_{B_I}\pa$) commute among each other,
and that the commutator of $\pa_{B_I}$ and $_{B_J}\pa$ is a combination of
$SO(1,1)$ and $SO(3,n-1)$ rotations, with moment maps given by
\be
h= \sqrt{\eta^{IJ} \eta_I\, \mu_J}=\hf\(\talp-\chi^A\tchi_A\) ,
\qquad
h_{IJ} = (\eta_I \mu_J - \eta_J \mu_I)/h\, .
\label{hhij}
\ee
Moreover, the moment maps $\eta^I, \mu_I$ transform covariantly under $SO(3,n-1)$ rotations
\be
\{ h_{IJ}, \eta_K \} =  -\eta_{IK} \eta_J + \eta_{JK}  \eta_I\, ,
\qquad
\{ h_{IJ}, \mu_K \} = -\eta_{IK} \mu_J + \eta_{JK}  \mu_I\, ,
\ee
and carry weight $+1,-1$ under $SO(1,1)$.

More importantly yet, the moment maps for the isometries $\pa_{B_I}$ and $_{B_I}\pa$
satisfy the quadratic constraints
\be
\label{quadcons}
\eta^{IJ} \eta_I\, \eta_J = 0,
\qquad
\eta^{IJ} \mu_I\, \mu_J = 0 ,
\ee
which reflect the fact that the Swann bundle $\cS$ of the orthogonal Wolf space
$\cW(n)$ is isomorphic to the minimal nilpotent orbit of $SO(n+4,\IC)$ \cite{MR1859424}.
Finally, the contact one-form \eqref{cXdarbA} can be expressed in terms of these moment maps as
\be
\label{cXdarbC}
\cX =  ( \eta^I \de\mu_I - \mu_I \, \de \eta^I) / h\, ,
\ee
which shows that $\eta_I/\sqrt{h},\mu_I/\sqrt{h}$ can be viewed as complex Darboux coordinates on $\cZ$,
despite being subject to the constraints \eqref{quadcons}.
For brevity, we shall refer to $\eta^I,\mu_I$ as the `heterotic moment maps'.

Although the moment maps defined above look complicated when expressed
in terms of the type II coordinates on $\cM_{\rm H}$, they acquire a very elegant form
in terms of the heterotic variables using the duality map \eqref{mirmaptree}.
Recalling the definition of the set of vectors $\gamma_I^x$ in \eqref{defgIx} and \eqref{param-zetatree},
we find
\be
\begin{split}
\eta_I = &e^{-R/2}\left( \gamma_I / t + \gamma_I^3 - \bar \gamma_I t \right) ,
\\
\mu_I =&
\( e^{-\rho}-\hf\,\eta^{JK}B_J B_K\)\eta_I +B_IB_J\eta^J+ e^{-\rho/2}\vartheta_{IJ}  B^J ,
\end{split}
\label{result-etamu}
\ee
where $\vartheta_{IJ}$  is the $\cO(2)$-tensor made out of the inner products
$\epsilon_{xyz} \gamma_I^y \gamma_J^z$,
\be
\vartheta_{IJ}=\I\, e^{-R/2}\[\(\gamma_I^3\gamma_J-\gamma_I\gamma^3_J\)/t
+2\(\gamma_I\bar\gamma_J-\bar\gamma_I\gamma_J\)+\(\gamma_I^3\bar\gamma_J-\bar\gamma_I\gamma^3_J\)t \] .
\label{defvarthet}
\ee
Likewise, the moment maps \eqref{hhij} for the $SO(1,1)$ and $SO(3,n-1)$ generators are
given by
\be
h=-B_I\eta^I \, ,
\qquad
h_{IJ}=e^{-\rho/2}\vartheta_{IJ} + B_I\eta_J-B_J\eta_I .
\label{generators2}
\ee
In particular, it is straightforward to check that under infinitesimal shifts  $B_I\to B_I+\eps_I$,
these generators shift according to the contact Poisson action \eqref{contaction}
with $\cO=\eps_I\eta^I$. More generally, under arbitrary finite shifts $B_I\to B_I+\eps_I$, the heterotic
moment maps transform according to
\be
\label{Bshift}
\eta_I\mapsto \eta_I ,
\qquad
\mu_I \mapsto \mu_I + (h_{IJ} - h \, \eta_{IJ}) \eps^J + \eps_I \eps^J \eta_J - \frac12\, \eps_J \eps^J \eta_I  .
\ee
Equations \eqref{quadcons}--\eqref{Bshift} provide an alternative
parametrization of the twistor lines of the
twistor space $\cZ\to \cW(n)$  which keeps symmetries under $SO(1,1)\times SO(3,n-1)$ rotations
and shifts $B_I\to B_I+\eps_I$ manifest.\footnote{Invariance under
$SO(3,n-1)$ is broken by the condition $\eta^1=1$ in \eqref{momapA}, but as we explain
in \S\ref{subsec-Swann}, this condition is an artifact of working on the twistor space $\cZ$
rather than the Swann bundle $\cS$.} In the next subsection we provide
a more conceptual derivation of this description which has the advantage of making
the full  $SO(4,n)$ symmetry manifest.

\subsection{Heterotic \hk quotient construction}
\label{subsec-Swann}

As explained in \cite{MR872143,Anguelova:2002kd}, the Swann bundle $\cS$ of the
orthogonal Wolf space $\cW(n)$ can be obtained as the \hk quotient of $\cH=\IH^{n+4}$ under
$SU(2)$ at zero level. To describe this quotient, we
choose complex coordinates $z^M_+,z^M_-$ on $\cH$ with $M$ running over the
indices $\flat,I,\sharp$, such that the complex symplectic form on $\cH$ is
\be
\label{omcH}
\omega_{\cH}=\eta_{MN}\, \de z_-^M\,\wedge\, \de z_+^N\, ,
\ee
where  $\eta_{MN}$ is the signature $(4,n)$ quadratic form defined in \eqref{defetaMN}.
The group $SU(2)$ of interest acts by left-multiplication on each of the quaternions
\be
\begin{pmatrix} z_+^M & -\bar z_-^M \\ z_-^M & \bar z_+^M \end{pmatrix} ,
\ee
while $SU(2)_R$ acts by right-multiplication.
The HK quotient $\cS=\cH/\!/\!/SU(2)$ is obtained by imposing the F-term
conditions\footnote{The level must be chosen to zero in order for the HK quotient to be a cone.}
\be
z^M_+ \eta_{MN} z^N_+ = z^M_+ \eta_{MN} z^N_- =z^M_- \eta_{MN} z^N_- = 0\, ,
\label{Ftermcond}
\ee
the D-term conditions
\be
z^M_+ \eta_{MN} \bar z^N_+ =z^M_- \eta_{MN} \bar z^N_- ,
\qquad
z^M_- \eta_{MN} \bar z^N_+ = 0 ,
\ee
and modding out by the $SU(2)$ action. Equivalently, one may  impose only
the F-term conditions and mod out by the action of the complexified group $SL(2,\IC)$
(on the open subset where this action has no fixed point).
In particular, away from the locus $z_-^\flat z_+^\sharp - z_-^\sharp z_+^\flat=0$, one may act
with the $SL(2,\IC)$ rotation
\be
\begin{pmatrix} a & b \\ c & d \end{pmatrix} =
\frac{1}{\sqrt{ z_-^\flat z_+^\sharp - z_-^\sharp z_+^\flat}}
\begin{pmatrix} z_-^\flat & - z_+^\flat \\ - z_-^\sharp & z_+^\sharp \end{pmatrix},
\label{compens}
\ee
to reach the $SL(2,\IC)$ gauge
\be
z'^\sharp_- = 0,
\qquad
z'^\flat_+=0,
\qquad
z'^\sharp_+ = z'^\flat_- .
\label{firstset}
\ee
Then the remaining coordinates are given by
\be
z'^I_+=\frac{z_-^\flat z_+^I - z_+^\flat z_-^I}{\sqrt{ z_-^\flat z_+^\sharp - z_-^\sharp z_+^\flat}} \, ,
\quad
z'^I_-=\frac{-z_-^\sharp z_+^I + z_+^\sharp z_-^I}{\sqrt{ z_-^\flat z_+^\sharp - z_-^\sharp z_+^\flat}} \, ,
\quad
z'^\sharp_+=z'^\flat_- = \sqrt{ z_-^\flat z_+^\sharp - z_-^\sharp z_+^\flat}\, .
\label{comptrans}
\ee
In this gauge the F-term constraints become
\be
z'^I_+ \eta_{IJ} z'^J_+ = 0 \, ,
\qquad
z'^I_- \eta_{IJ} z'^J_- = 0\, ,
\qquad
z'^I_- \eta_{IJ} z'^J_+ =(z'^\sharp_+)^2\, ,
\label{secondset}
\ee
while the symplectic form \eqref{omcH} reduces to
\be
\label{omcS2}
\omega_{\cS}=\eta_{IJ}\, \de z'^I_-\,\wedge\, \de z'^J_+\, .
\ee

The moment maps for the triholomorphic action of $SO(4,n)$ on the quotient $\cS$ are simply
the restriction of the moment maps of the $SO(4,n)$ action on $\cH$ to the locus \eqref{Ftermcond}, namely
\be
\hh^{MN} = z^M_+ \, z^N_- -z^N_+\, z^M_- \, .
\label{GenHMN}
\ee
Identifying the moment maps $\hh^{I\flat},\hh^{\sharp I},\hh^{IJ},\hh^{\sharp\flat}$ on $\cS$
with the moment maps $\eta^I, \mu^I, h^{IJ}, h$ on $\cZ$, rescaled by the $\IC^\times$
coordinate $\nu$, we find
\be
z'^I_+ = \sqrt{\nu/h}\, \eta^I\, ,
\qquad
z'^I_- = \sqrt{\nu/h}\, \mu^I\, ,
\qquad
z'^\sharp_+ = z^\flat_- = \sqrt{\nu\, h}\, .
\label{zpmunu}
\ee
In particular, under these identifications  the quadratic conditions \eqref{secondset}
reproduce the conditions \eqref{quadcons} satisfied by the `heterotic moment maps',
as well as the first relation in \eqref{hhij}.
The coordinate $\nu$ may be set to 1 at the expense of relaxing the gauge condition $\eta^1=1$ in \eqref{momapA}.
Finally, one easily checks that substituting \eqref{zpmunu}  into \eqref{omcS2},
one reproduces the symplectic form \eqref{gensymform} with $\cX$ as in \eqref{cXdarbC}.

The identification \eqref{zpmunu} shows that the heterotic moment maps $\eta^I,\mu_I$ subject
to the constraints \eqref{quadcons}  are indeed (after an overall rescaling by $\sqrt{h}$)
Darboux coordinates on the Swann bundle $\cS$,
inherited from the unconstrained coordinates $z_\pm^M$ on the total space
$\cH=\mathbb{H}^{n+4}$ before quotienting by $SL(2,\IC)$. Unlike the type IIB (unconstrained)
Darboux coordinates $\txi^A,\xi_A,\talp,\nu$, the coordinates $\eta^I,\mu_I$
transform linearly under $SO(3,n-1)$, and according to \eqref{Bshift} under shifts of $B_I$.
As we shall see in the next Section, they provide a convenient framework
to describe quantum corrections on the heterotic side.

\section{Heterotic worldsheet instantons in twistor space}
\label{sec-wsinst}

In the previous section, we provided a new system of Darboux coordinates on the twistor space
(or Swann bundle) of the HM moduli space in the classical limit,
where it reduces to the orthogonal Wolf space $\cW(n)$. The main advantage of these new
coordinates $\eta^I,\mu_I$ is that they transform linearly under $SO(3,n-1)$, and have simple
transformation properties \eqref{Bshift} under large gauge transformations of the $B$-field.
The price to pay is that they are not independent, rather they satisfy
the invariant quadratic constraints \eqref{quadcons}. In this section, using this framework, we investigate the
twistorial representation of the worldsheet instanton corrections which are expected to correct
the HM moduli space away from the classical limit.
Perturbative $\alpha'$ corrections,
and a detailed comparison of quantum corrections on the heterotic and type II side is left for future work.

\subsection{Generalities on heterotic worldsheet instantons}

To all orders in a perturbative $\alpha'$ expansion, corrections to the HM moduli space
are expected to be independent of the $B$-field moduli $B_I$. In contrast, worldsheet instantons
wrapping a two-cycle $\tau=p^I\tau_I \in H_2(\Kf,\IZ)$  are characterized by an exponential
dependence of the form $e^{\mp 2\pi\I p^I B_I}$ on these moduli, coming from the topological
coupling $\int_{\tau} B$ of the heterotic string worldsheet to a background Kalb-Ramond field.
This dependence ensures that the moduli space is invariant under large gauge transformations
$B\mapsto B + \eps_I \omega^I$ of the $B$ field, where the transformation parameters $\eps_I$ are integers.

In the large volume limit, worldsheet instanton contributions  are expected to be of order
$e^{-2\pi S_{\rm inst}}$ where $S_{\rm inst}$ is the classical Nambu-Goto action of the
heterotic string wrapped on $\tau$ supplemented by the usual axionic coupling,
\be
S_{\rm inst} = e^{-\rho/2} \sqrt{ D } \pm \I  p^I B_I .
\label{instact}
\ee
Here $D$ is the square of the area of the supersymmetric cycle in the homology class $\tau$.
Since supersymmetric cycles are complex curves for any linear combination of the triplet
of complex structures $\cJ^x$ on $\cS$, their area is given by
\be
D=(p^I\gamma_I^x)(p^J\gamma_J^y)\delta_{xy}=p^I (\Mi_{IJ} + \eta_{IJ} ) p^J .
\label{defD}
\ee
Moreover, since the HM moduli space should be independent of the heterotic dilaton, corrections
are only expected to come at tree-level, i.e. for wordsheets of genus 0. This constrains the
homology class $\tau$ to satisfy
\be
p^I \eta_{IJ} p^J = 2g-2 = -2\, .
\label{requi}
\ee

The prefactor in front of the exponential $e^{-2\pi S_{\rm inst}}$
 should include the one-loop fluctuation determinant
around the classical configuration and other factors which depend
on the observable of interest (for example, a component of the metric tensor). It is expected
to be uniquely determined by supersymmetry constraints, up to a factor
$\Omega_{\rm H}(p^I)$ which should be a topological
invariant of the K3 surface $\Kf$ and $E_8\times E_8$ gauge bundle (in particular,
locally independent of all moduli). Unlike the D-instanton
corrections on the type IIB side, this topological invariant does not appear to have been
formulated in the mathematical literature. It  would be interesting to investigate whether
$\Omega_{\rm H}(p^I)$  is liable to jump at certain loci in HM moduli space, and whether
it may be subject to certain global consistency conditions. In this work we shall work
at linear order in the instanton corrections, and treat $\Omega_{\rm H}(p^I)$ as unknown constant parameters.

\subsection{A twistorial origin for scalar-valued worldsheet instantons\label{sec_scal}}

As recalled in  section \ref{subsec-twistor}, linear perturbations of the classical metric on
$\cM$ which preserve the QK property are described  by elements of $H^1(\cZ,\cO(2))$ ---
in practice, by a set of holomorphic functions $\Hij{ij}_{(1)}$ on the overlaps of two patches
in an open covering $\cU_i$ of $\cZ$. Viewed as homogeneous functions on the Swann bundle $\cS$,
these must transform with degree 1 under the action of $\IC^\times$ \cite{Alexandrov:2008nk}.

Rather than studying these quaternionic metric fluctuations,
we shall discuss a much simpler problem involving scalar-valued functions $\Psi$ on $\cM$,
whose twistorial description is very
similar to the one relevant for metric perturbations. Namely, we consider zero-modes of the
second-order differential operator (sometimes known as the Baston operator)
\cite{MR1165872}\footnote{This vector-valued operator reduces to
the conformal Laplacian in quaternionic dimension $d=1$. We refer to \cite{Neitzke:2007ke} for notations and details on this operator.}
\be
\label{Baston}
\left[ \nabla_{AA'} \nabla^{A'}_{\,B} - \frac{\eps_{AB}}{4d(d-2)}\, R \right] \Psi = 0\, ,
\ee
where
$\nabla_{AA'}$ is the $Sp(d)\times SU(2)_R $ spin connection on $\cM$ and $R$ is
the constant scalar curvature on $\cM$.
It is known that the solutions of \eqref{Baston} are in
one-to-one correspondence with elements of $H^1(\cZ,\cO(-2))$ via the Penrose
transform \cite{MR1165872,Neitzke:2007ke}. The latter can again be represented by
a set of holomorphic functions $\Hij{ij}_{(-1)}$ on the overlaps $\cU_i\cap \cU_j$
which now descend from homogeneous functions of degree $-1$ on the Swann bundle.
The solution to \eqref{Baston} follows  by a simple Penrose-type contour integral
\be
\Psi = \sum_j\oint_{C_j}  \nu\cX\, \Hij{ij}_{(-1)} \sim
r \sum_j\oint_{C_j} \frac{\de t}{t}\, \nu\Hij{ij}_{(-1)}  ,
\ee
where $\cX$ is the contact one-form \eqref{XDt}, $r=e^{\Phi}=\hf\,e^{-(\rho+R)/2}$ is the contact potential,
$\nu$ is the projective coordinate which can be absorbed due to the homogeneity of $\Hij{ij}_{(-1)}$,
and $C_j$ is a contour on $S^2$ surrounding the patch
$\cU_j$.\footnote{As in \cite{Alexandrov:2008ds,Alexandrov:2008nk}, the sum over patches
does not depend on the free index $i$ due to the standard consistency conditions on
the holomorphic functions  $\Hij{ij}_{(-1)}$.}
Our goal is to identify
a holomorphic function $H_{(-1)}$ and the associated contour $C$ which, after Penrose transform,
produces a function $\Psi$ on $\cM$ which behaves
as  $e^{-2\pi S_{\rm inst}}$ in the large volume limit.

Our starting point is the observation that shifts $B_I\mapsto B_I+\eps_I$
have an especially simple action on the space $\cH=\IH^{n+4}$ prior to
the $SL(2,\IC)$ quotient,\footnote{The transformation \eqref{BshiftH} can be obtained by acting
with the matrix appearing in \eqref{MatrixM} where $B_I$ is replaced by $\epsilon_I$.
Note that it does not preserve the $SL(2,\IC)$ gauge \eqref{firstset}. However it does reduce
to \eqref{Bshift}, when followed by an $SL(2,\IC)$ rotation to restore \eqref{firstset}.
See appendix \ref{ap-quotient} for details.}
\be
\label{BshiftH}
z_\pm^\flat \mapsto z_\pm^\flat ,
\qquad
z_\pm^I\mapsto z_\pm^I - \eps^I z_\pm^\flat ,
\qquad
z_\pm^\sharp \mapsto z_\pm^\sharp - \eps_I z_\pm^I
+ \frac12\, \eps_I \eta^{IJ} \eps_J\, z_\pm^\flat .
\ee
In particular, for any integer $m$ the function on $\cH$
\be
\hat H(\zp^N)=(\zp^\flat)^{2m} e^{2\pi \I p_I \zp^I/\zp^\flat}
\ee
is invariant under $B$-shifts with $\epsilon_I$ integer. However, it is not invariant
under $SL(2,\IC)$, so does not descend to the quotient $\cS$.
To repair this problem, one may formally average over the action
of the nilpotent  subgroup of $SL(2,\IC)$
\be
\(\begin{array}{c} \zp^N \\ \zm^N \end{array}\)\mapsto
\(\begin{array}{cc} 1 & b \\ 0 & 1 \end{array}\)
\(\begin{array}{c} \zp^N \\ \zm^N \end{array}\)
\ee
to obtain an invariant function
\be
\label{averag}
\begin{split}
\int \de b\,\hat H(\zp^N+b\zm^N)= &\,
\int \de b\,\hat H(\zp'^N+b\zm'^N)
=\,\int \de b\, (b\sqrt{\nu h})^{2m} e^{2\pi\I p_I \(z'^I_- +\frac{1}{b}\,z'^I_+\)/\sqrt{h}}
\\
\sim &\, \nu^{m} h^{-1/2}(p_I z'^I_+)^{2m+1} e^{2\pi\I p_I z'^I_-/\sqrt{h}}
=\,\nu^{m} h^{-m-1}(p_I\eta^I)^{2m+1}e^{2\pi\I p_I\mu^I/h},
\end{split}
\ee
where we used the invariance of the Haar measure and the gauge conditions \eqref{firstset}.
This formal computation works for any $m$, but the requirement of degree $-1$ weight
under the $\IC^\times$ action fixes $m=-1$.
This motivates the following Ansatz:
\be
\label{H1Om2}
\nu H_{(-1)} = \frac{\Omega_{\rm H}(p^I)}{p^I \eta_I} \,
\exp\left( \pm 2 \pi \I\, p^I \mu_I / h \right) ,
\ee
where the  contour $C$ should be chosen to encircle one of the roots of
the denominator $p^I \eta_I(t)$,
\be
\label{tpm}
t_\pm=\frac{p^I\gamma^3_I \pm \sqrt{D}}{2 p^J\bar\gamma_J}\, .
\ee
An explicit computation given in appendix \ref{ap-instact}
shows that in this case the Penrose transform of \eqref{H1Om2} is
\be
\label{PsiPen}
\Psi \sim \frac{e^{-\rho/2}}{\sqrt{D}} \, \Omega_{\rm H}(p^I)\,
\exp\left( -2\pi e^{-\rho/2} \sqrt{D} \mp 2\pi\I p^I B_I \right),
\ee
reproducing the expected classical action \eqref{instact}.

It is important to note that, before the Penrose transform, the holomorphic function \eqref{H1Om2}
is not invariant under integer shifts of $B_I$. Instead it transforms
by a term which is regular at the zeros of $p^I \eta_I$ and therefore does not contribute
to the contour integral. Another important observation is that
wave-functions of the precise form \eqref{PsiPen} appear as Fourier coefficients of certain
automorphic forms  of $SO(4,n,\IZ)$ obtained by the so-called
Borcherds' (or theta) lift. We explain this relation and comment
on its possible physical implications in Appendix \ref{sec_theta}.

\subsection{Generating functions for heterotic symplectomorphisms}
\label{subsec-genfun}

We now return to the question of worldsheet instanton corrections to the classical HM moduli space.
At linear order, these corrections correspond to linear perturbations of the homogenous
complex symplectomorphisms relating Darboux coordinate systems on
the Swann bundle $\cS$. As in the previous subsection, it would be desirable to parameterize
these infinitesimal symplectomorphisms  by holomorphic functions $H(\eta^I,\mu_I)$
of {\it unconstrained} Darboux coordinates $\eta^I,\mu_I$, as these variables
transform linearly under $SO(3,n-1)$ rotations and shifts of the $B$-field.
At the same time, one must ensure that the resulting transformations preserve
the quadratic constraints \eqref{quadcons} satisfied by
the heterotic moment maps.

One way to achieve such a parameterization is i) to consider the most general infinitesimal
symplectomorphism on the flat HK space $\cH$ prior to quotienting
under $SL(2,\IC)$, ii) require that it preserves the F-term conditions \eqref{Ftermcond},
and iii) compose it with an $SL(2,\IC)$ rotation so as to maintain the $SL(2,\IC)$ gauge choice \eqref{firstset}.
As explained in Appendix \ref{ap-quotient}, one finds that the most general
infinitesimal transformation of the  heterotic moment maps $\eta^I,\mu_I$
preserving the symplectic form \eqref{omcS2} in the gauge $\nu=1$,
\be
\label{omcS3}
\omega_{\cS}= \de\(\mu_I/\sqrt{h}\)\,\wedge\, \de\(\eta^I/\sqrt{h}\),
\qquad
h=\sqrt{\eta^I\mu_I}
\ee
and the quadratic constraints \eqref{quadcons} is given by
\be
\delta \eta^I=\(h^{IJ}+h\, \eta^{IJ}\)\p_{\mu^J}H,
\qquad
\delta \mu^I=\(h^{IJ}-h\, \eta^{IJ}\)\p_{\eta^J}H,
\label{holaction}
\ee
where $H$ is an arbitrary holomorphic function of the heterotic moment maps
$\eta^I,\mu_I$ (viewed as unconstrained coordinates), and $h^{IJ}, h$ are defined in \eqref{hhij}. In order that
this transformation commutes with the $\IC^\times$ action, it is furthermore
necessary that $H$ be a homogeneous function of degree 1 in $\eta^I,\mu_I$.

In order to describe worldsheet instanton corrections, we must require that $H(\eta^I,\mu_I)$
be invariant under integer shifts of $B_I$, up to regular terms. Unfortunately, the
naive Ansatz \eqref{averag} with $m=1$ (so as to have the proper weight under $\IC^\times$ action) fails,
as the resulting prefactor $(p_I \eta^I)^3/h^2$ in front of the exponential is no longer singular
at the zeros $t_\pm$ of $p_I \eta^I$. We leave it as an open problem to determine the correct
function $H(\eta^I,\mu_I)$ describing worldsheet instanton corrections to the HM moduli space
at linear order, or even more ambitiously at the non-linear level.

\section{Discussion \label{sec_disc}}

In this note we reconsidered the duality between the $E_8\times E_8$ heterotic string compactified
on $\Kf\times T^2$, where $\Kf$ is an elliptically fibered K3 surface,
and the type IIB string compactified on a $K3$-fibered CY threefold $\CYm$, from
the point of view of their hypermultiplet moduli spaces. In the classical limit
where both $\Kf$ and $\CYm$ are large (and so are the bases of the respective fibrations),
and assuming the absence of bundle moduli on the heterotic side,
or reducible singular fibers on the type IIB side, the HM moduli spaces are
locally isomorphic to the orthogonal Wolf space $\cW(n)$.
We presented a simple relation \eqref{mirmaptree} between
the natural type IIB and heterotic fields which identifies the two metrics,
improving the duality map of  \cite{Louis:2011aa} so as to match the global identifications under
large gauge transformations on both sides.

More importantly, we introduced a new
twistorial description of the HM moduli space which manifestly exhibits
the symmetries of the heterotic string, including in particular
the large gauge transformations of the B-field moduli and
automorphisms of the homology lattice $H_2(\Kf,\IZ)$.
We related this description to the
standard twistorial formulation of the HM moduli space on the type IIB side, inherited from
the $c$-map construction \cite{Neitzke:2007ke,Alexandrov:2008nk,Alexandrov:2008gh}.
It would be very interesting to relate this description to the pure spinor formalism
for heterotic compactifications developed e.g. in \cite{Chandia:2011su}.

Going beyond the classical limit, we started investigating twistorial
description of heterotic worldsheet instantons,
at the level of linear perturbations around the classical moduli space $\cW(n)$.
While we did not reach
a conclusive answer to this problem, we found a simple holomorphic function \eqref{H1Om2}
which, upon Penrose transform, reproduces the zero-modes of the second-order
Baston operator \eqref{Baston} with exponential dependence on the $B$-field moduli.
This provides
an important clue on the twistorial lift of genuine heterotic worldsheet instantons.

Clearly, much remains to be done before establishing heterotic-type II duality
in the hypermultiplet sector. First, one should extend the duality map  to include
gauge bundle moduli on the heterotic side. While the contribution of  reducible singular
fibers on the type IIB side is computable using mirror symmetry, a set of canonical coordinates
on the gauge bundle moduli space is so far lacking, making this extension difficult.
It may be possible to improve this situation for bundles admitting a spectral
cover construction \cite{Friedman:1997yq}, where the the HK metric on the
gauge bundle moduli space is expected to be the rigid $c$-map metric
on the Jacobian of the spectral curve.

Another interesting problem is to match
the 4-loop $\alpha'$ correction to the type IIB prepotential and
the 1-loop $g_s$ correction to the $c$-map metric \eqref{hypmettree}
(both of which proportional to the Euler number of $\CYm$
\cite{Antoniadis:1997eg,Gunther:1998sc,Antoniadis:2003sw,Robles-Llana:2006ez})
against some perturbative corrections on the heterotic side.
While both effects will change
the form of the twistor lines (i.e.  the expressions \eqref{gentwi}, \eqref{result-etamu}
of the Darboux coordinates in terms of coordinates on $\cM\times S^2$), it is worth noting that
 neither the relations \eqref{momapA}, \eqref{momapB} between the type IIB Darboux coordinates
$(\xi^\Lambda,\txi_\Lambda,\talp)$ and the constrained heterotic Darboux coordinates
$(\eta^I,\mu_I)$,  nor the quadratic constraints \eqref{quadcons} themselves, will be affected.
As on the type II side, these loop corrections are crucial for establishing the correct
topology of the perturbative moduli space, and impose important constraints
on non-perturbative corrections \cite{Alexandrov:2010np,Alexandrov:2010ca}. On the heterotic side
these loop corrections  will in general spoil the flatness of the torus bundle of
$B$-field moduli $B_I$ over the moduli space of metrics on K3, and should cancel against
 chiral anomalies on the heterotic string worldsheet, such that
worldsheet instanton corrections are well-defined \cite{Witten:1999fq}.

Having matched perturbative contributions, one should then be able to compare the worldsheet,
D-brane and NS5-brane instantons on the type IIB side with worldsheet instanton corrections
on the heterotic side. This of course requires a full understanding of these heterotic
instantons in twistor space, which we have achieved only partially. One intriguing issue is that,
on the face of Eq. \eqref{Bmap},
both type IIB worldsheet instantons wrapping the base $\CP$ of the K3-fibration, and NS5-branes
wrapping the full CY threefold, seem to map to worldsheet instantons wrapping
the genus-one curves $\cE$ and $\cE+\cB$. This is in apparent contradiction
with the fact that the metric on the HM moduli space should
be exact at tree-level and thus receive corrections from  genus-zero worldsheet instantons only.
We expect that this puzzle will be resolved by including bundle moduli,
which could modify the duality map and
relax the constraint \eqref{requi}.
Assuming that these issues can be cleared, one may hope that heterotic-type II duality,
along with other dualities between $N=2$ string vacua, will
provide new tools for computing the exact metric on the hypermultiplet moduli space, and
extracting the  topological invariants encoded in instantonic effects.

As an interesting spin-off of our study of worldsheet instantons in twistor space, we have
found in Appendix \ref{sec_theta} a twistorial description of a class of automorphic forms under
$SO(4,n,\IZ)$ obtained from a weakly holomorphic modular form of weight $(4-n)/2$
by Borcherds' lift, whose Fourier coefficients have the expected form
of heterotic worldsheet instantons. Unfortunately,
the twistorial description of these scalar-valued automorphic forms is in terms
of $H^1(\cZ,\cO(-2))$, rather than $H^1(\cZ,\cO(2))$ as required for
worldsheet instanton corrections to the QK metric. It would be very interesting
to identify a class of heterotic/type II duals where a discrete
subgroup $SO(4,n,\IZ)$ of the isometry group of the moduli space $\cM_{\rm H}$
could arguably stay unbroken at the non-perturbative level in the spirit of
\cite{RoblesLlana:2006is,Pioline:2009qt,Bao:2009fg}.
Determining the automorphic form in $H^1(\cZ,\cO(2))$ governing these quantum corrections
could then provide an alternative way to arrive at the afore-mentioned topological invariants.

\acknowledgments

It is a pleasure to thank J. Louis and R. Valandro for valuable discussions on heterotic/type II duality.
B.P. is also grateful to S. Hohenegger and D. Persson for discussions on automorphic forms,
Borcherds' lift and quaternionic discrete series.

\appendix

\section{$SU(2)$ quotient and symplectomorphisms}
\label{ap-quotient}

Infinitesimal symplectomorphisms on the flat HK space $\cH=\IH^{n+4}$ are given by the following transformations
\be
\delta \zp^M=\eta^{MN}\,\frac{\p \hH}{\p \zm^N},
\qquad
\delta \zm^M=-\eta^{MN}\,\frac{\p \hH}{\p \zp^N},
\label{bigsymplect}
\ee
where $\hH(\zp,\zm)$ is an arbitrary holomorphic function assumed to be infinitesimally small.
We are interested in describing the symplectomorphisms which descend to the \hk quotient
$\cH/\!/\!/ SU(2)$.
This amounts to imposing that the transformation should preserve the F-term conditions \eqref{Ftermcond}
and the gauge conditions \eqref{firstset}.

The first restriction can easily be satisfied. Indeed, since the F-term conditions are
invariant under $SO(4,n)$,
they are preserved if and only if the generating function $\hH$ depends on $\zpm^M$ only
through the $SO(4,n)$ generators  $\hh^{MN}$ \eqref{GenHMN}.
From \eqref{bigsymplect}, one then obtains
\be
\begin{array}{rclrcl}
\delta \zp^\flat &=& \zp^J \p_{h^{\sharp J}}\hH,
&\qquad
\delta \zm^\flat &=& \(\zm^J \p_{h^{\sharp J}}+\sqrt{h}\p_{h^{\sharp\flat}}\)\hH,
\\
\delta \zp^\sharp &=& -\(\zp^J \p_{h^{J\flat}}+\sqrt{h}\p_{h^{\sharp\flat}}\)\hH,
&\qquad
\delta \zm^\sharp &=& -\zm^J \p_{h^{J\flat}}\hH,
\\
\delta \zp^I &=&-\eta^{IJ}\( 2\zp^K \p_{h^{JK}} -\sqrt{h}\p_{h^{\sharp J}}\)\hH,
&\qquad
\delta \zm^I &=& -\eta^{IJ}\(2\zm^K \p_{h^{JK}} +\sqrt{h}\p_{h^{J \flat}}\)\hH,
\end{array}
\label{trans-hhh}
\ee
where we assumed that $\zpm^M$ satisfy the gauge conditions \eqref{firstset}.
It is trivial to check that \eqref{trans-hhh} does preserve \eqref{Ftermcond}, but
it spoils the gauge conditions  \eqref{firstset}. To restore these conditions, we need to supplement
the symplectomorphism \eqref{trans-hhh} by a compensating gauge transformation \eqref{compens}.
Expanding this transformation in variations of $\zpm^M$, one finds
\be
\begin{split}
\delta z'^I_+ = &\, \delta z^I_+ +\frac{z^I_+}{2\sqrt{h}}\(\delta z_-^\flat-\delta z_+^\sharp\)
- \frac{z^I_-}{\sqrt{h}}\, \delta z_+^\flat,
\\
\delta z'^I_- = &\, \delta z^I_- +\frac{z^I_-}{2\sqrt{h}}\(\delta z_+^\sharp-\delta z_-^\flat\)
- \frac{z^I_+}{\sqrt{h}}\, \delta z_-^\sharp.
\end{split}
\label{varzp}
\ee
Finally, the relations \eqref{zpmunu} (where we take $\nu=1$ thereby relaxing the gauge $\eta^1=1$) imply
\be
\begin{split}
\delta \eta^I = &\, \sqrt{h} \delta z'^I_+ +\frac{z'^I_+}{2\sqrt{h}}\, \eta_{JK}\(z'^J_+\delta z'^K_- +z'^J_-\delta z'^K_+\),
\\
\delta \mu^I = &\, \sqrt{h} \delta z'^I_- +\frac{z'^I_-}{2\sqrt{h}}\, \eta_{JK}\(z'^J_+\delta z'^K_- +z'^J_-\delta z'^K_+\).
\end{split}
\label{varetamu}
\ee
Substituting \eqref{trans-hhh} into \eqref{varzp} and the latter into \eqref{varetamu}, one can show that
\be
\begin{split}
\delta \eta^I = &\, -2\eta^J\p_{h^{IJ}}\hH +\(h^{IJ}+h\eta^{IJ}\)\p_{h^{\sharp J}}\hH+\eta^I\p_{h^{\sharp\flat}}\hH,
\\
\delta \mu^I = &\,  -2\mu^J\p_{h^{IJ}}\hH +\(h^{IJ}-h\eta^{IJ}\)\p_{h^{J\flat}}\hH-\mu^I\p_{h^{\sharp\flat}}\hH,
\end{split}
\ee
On the other hand, one arrives at exactly the same result if one takes the holomorphic function in \eqref{holaction}
to be given by $H(\eta^I,\mu_I)=\hH(\hh^{MN})$ where the generators are expressed through the generalized Darboux coordinates
by means of \eqref{zpmunu}.

\section{Penrose transform}
\label{ap-instact}

In this appendix we evaluate of the Penrose transform of \eqref{H1Om2}.  The contour
integral picks up the residue of this function at one of the zeros $t=t_\pm$ of
$p^I \eta_I=0$. To compute the residue we need to evaluate the argument of the exponential
$p^I\mu_I/h$ at the same point.  From \eqref{result-etamu} and \eqref{generators2}, one finds
\be
\left.\frac{p^I \mu_I}{h}\right|_{p^I\eta_I=0} = - p^I B_I - e^{-\rho/2}\, \frac{p^I  \vartheta_{IJ}B^J}{\eta_K B^K} \, .
\label{Sinsttwistor}
\ee
To  evaluate the second term, let us make some preliminary observations.
Recall that an $\cO(2)$ section of the form $\eta(t)=a/t+b-\bar a t$
can be associated to a vector $\vec \eta = (a,b,\bar a)$ in $\IR^3$ equipped with
the Euclidean metric $\|\vec \eta\|^2=b^2 + 4 |a|^2$. Given two vectors $\vec\eta$ and $\vec\eta'$,
the inner product
$\vec \eta'' = \vec \eta \wedge \vec \eta'
= \I ( a' b-ab' , 2(a\bar a'-a' a), \bar a b'- b \bar a')$
corresponds to the $\cO(2)$ section
\be
\vec \eta'' = \I \left( \frac{a' b-ab'}{t} + 2(a\bar a'-a' \bar a) + t (\bar a' b - \bar a b') \right) .
\ee
On the other hand, the twistor coordinate $t$ parametrizes a unit vector
$\vec t=(-t,1-|t|^2,-\bar t)/(1+|t|^2)$. A simple computation shows that
\be
| \eta(t) |^2 = \frac{1+|t|^2}{4|t|^2}\, \| \vec\eta \wedge \vec t\|^2\, .
\ee
In particular, this implies that $\eta(t)$ vanishes for $\vec t=\pm \vec\eta/ \|\vec\eta\|$.
Using these remarks, we can recast \eqref{result-etamu} and \eqref{defvarthet} as
\be
\vec \eta_I = e^{-R/2} \vec \gamma_I\, ,
\qquad
\vec \vartheta_{IJ}=e^{-R/2} \vec \gamma_I \wedge \vec \gamma_J\, .
\ee

Returning to \eqref{Sinsttwistor}, we note that the second term is pure imaginary.
The norm of the numerator can be evaluated as follows
\be
\begin{split}
| p^I \vartheta_{IJ}(t) B^J |^2 = &
\frac{(1+|t|^2) e^{-R}}{4|t|^2 \| p^L \vec \gamma_L \|^2}\,
\| (p^I \vec \gamma_I) \wedge (p^J \vec \gamma_J) \wedge (B^K \vec \gamma_K)  \|^2
\\
=& \frac{(1+|t|^2) e^{-R}}{4|t|^2 \| p^L \vec \gamma_L \|^2}\, \| p^I \vec \gamma_I\|^2\,
\|(p^J \vec \gamma_J) \wedge (B^K \vec \gamma_K)  \|^2,
\end{split}
\ee
where we used the fact that, for two vectors in $\IR^3$,
$\| \vec a\wedge (\vec a\wedge \vec b)\|^2=\| \vec a\|^2 \, \| \vec a\wedge \vec b \|^2$.
On the other hand, the norm of the denominator is given by
\be
|h(t)|^2= |B^I \eta_I(t)|^2 =
\frac{(1+|t|^2) e^{-R}}{4|t|^2 \| p^L \vec \gamma_L \|^2}\, \|(p^J \vec \gamma_J) \wedge (B^K \vec \gamma_K)  \|^2
\ee
so that the ratio is found to be
\be
\frac{p^I \vartheta_{IJ}(t) B^J}{h(t)} = \pm \I \| p^I \vec \gamma_I \|\, .
\ee
As a result, we indeed recover the worldsheet instanton action \eqref{instact}
\be
\left.\frac{p^I \mu_I}{h}\right|_{p^I\eta_I=0} =\pm \I S_{\rm inst}
\ee
and the formula \eqref{PsiPen} immediately follows.

\section{Twistorial theta lift \label{sec_theta}}

In this Appendix we expand on the observation mentioned in \S\ref{sec_scal}
that the eigenmodes \eqref{PsiPen} of the Baston differential operator \eqref{Baston}
arise as Fourier coefficients of certain automorphic forms  obtained by the theta lift procedure.
For simplicity, we assume that $n-4$ is a multiple of 8.\footnote{If not, one must relax
the assumption that the lattice $\Lambda$ is self-dual and replace the modular group
$SL(2,\IZ)$ appearing below  by one of its congruence subgroups. The general conclusion
however still holds.}
Let $\Lambda$ be an even self-dual Lorentzian lattice with signature $(4,n)$.
The Narain lattice partition function
\be
Z_{4,n}(\tau) = \tau_2^2\, \sum_{p^M \in \Lambda}
e^{-\pi \tau_2\, p^M \Mn_{MN} p^N -\pi\I \tau_1 p^M \eta_{MN} p^N} ,
\label{partfun}
\ee
is a function on the Poincar\'e upper half plane, parametrized by $\tau=\tau_1+\I\tau_2$,
times the orthogonal Wolf space $\cW(n)$,  which we parametrize by the
same matrix $\Mn_{MN}$ as in \eqref{MatrixM}.
It is a modular form of weight $((n-4)/2,0)$ under the usual fractional linear action
of $SL(2,\IZ)$ on $\tau$, invariant under the automorphism group $SO(4,n,\IZ)$  of the lattice $\Lambda$.
Given any weakly holomorphic modular form $\Phi(\tau)$ of weight $(4-n)/2$, the (suitably regularized) modular integral
 \be
\label{defLift}
\Psi =  \int_{\cF} \frac{\de\tau_1\de\tau_2}{\tau_2^2}\,
\Phi(\tau)\, Z_{4,n}(\tau)
\ee
over a fundamental domain $\cF$ of $SL(2,\IZ)$ gives a function on $\cW(n)$,
invariant under the automorphism group $SO(4,n,\IZ)$, which is
known as the theta lift or Borcherds' lift of the modular form $\Phi$
 \cite{0919.11036}. In particular, using the same
parametrization of  $\cW(n)$ as in the body of this paper,
$\Psi$ is invariant under integer shifts $B_I\mapsto B_I+\eps_I$.

The Fourier coefficients of $\Psi$ with respect to
the periodic variables $B_I$ can be computed by using a Poincar\'e series representation for
the lattice partition function, obtained by Poisson resummation with respect to $p^\flat$ in \eqref{partfun}.
Denoting by $k$ the integer dual to $p^\flat$ and $w=p^\sharp$,
this gives
\be
Z_{4,n}(\tau) = e^{-\rho/2}\, \sum_{k,w}
 \sum_{\tilde p^I \in \Gamma_{3,n-1}+w B^I} \!\!\!\! \tau_2^{3/2}
 e^{-\pi \tau_2\, \tilde p^I \Mi_{IJ} \tilde p^J
-\pi\I \tau_1 \tilde p^I \eta_{IJ} \tilde p^J
 -\frac{\pi}{\tau_2}\, e^{-\rho} |k-w\tau|^2+2\pi\I k B_I \tilde p^I-\I\pi kw B^2}\, .
\ee
 Since $SL(2,\IZ)$ acts linearly on $(k,w)$, the sum over
$(k,w)\neq(0,0)$ can be restricted to $k\neq 0$, $w=0$, at the expense
of extending the integration domain $\cF$ to the strip $\{-1/2\leq \tau_1<1/2,\ \tau_2>0\}$.
Fourier expanding $\Phi(\tau)=\sum_{N\geq \kappa} c(N) e^{2\pi\I N \tau}$, we find
\beq
\label{psi2}
\Psi &=&
 e^{-\rho/2}\, \int_{\cF} \frac{\de\tau_1\de\tau_2}{\tau_2^2}\,
\Phi(\tau) \, \sum_{\tilde p^I \in \Gamma_{3,n-1}} \!\!\!\! \tau_2^{3/2}
 e^{-\pi \tau_2\, \tilde p^I \Mi_{IJ} \tilde p^J
-\pi\I \tau_1 \tilde p^I \eta_{IJ} \tilde p^J}
\\
 &+& e^{-\rho/2} \int_0^{\infty} \int_{-1/2}^{1/2}  \frac{\de\tau_1\de\tau_2}{\tau_2^{1/2}}
\sum_{\tilde p^I\in \Gamma_{3,n-1}}  \sum_{N\geq \kappa} \sum_{k\neq 0}
 c(N)\, e^{-\pi \tau_2\,\tilde p^I \Mi_{IJ} \tilde p^J
-\pi\I \tau_1 \tilde p^I \eta_{IJ} \tilde p^J +2\pi\I N\tau -\frac{\pi }{\tau_2}\,e^{-\rho} k^2+2\pi\I k B_I \tilde p^I}\, .
\nn
\eeq
The first term is independent of the variables $B_I$ and automorphic under $SO(3,n-1,\IZ)$.
From the   second term one can read off the Fourier coefficient of $\Psi$
with non-zero momentum $p^I=k \tilde p^I$ with respect to $B_I$,
\be
\Psi(p^I) = e^{-\rho/2} \,
\Omega_{\rm H}(p^I)\,
\frac{\exp\left( -2\pi e^{-\rho/2} \sqrt{p^I (\Mi_{IJ}+\eta_{IJ} )p^J}
+ 2\pi\I p^I B_I \right) }{\sqrt{p^I (\Mi_{IJ}+\eta_{IJ} )p^J} }\, ,
\label{resPsi}
\ee
where we defined
\be
\label{OmHc}
\Omega_{\rm H}(p^I) = \sum_{k|p^I}
k\, c\( p^I \eta_{IJ} p^J/(2k^2)\)\, .
\ee
The resulting function \eqref{resPsi} has precisely the same form as the
wave-function \eqref{PsiPen}. Note that \eqref{OmHc} vanishes whenever
$p^I$ is primitive and $p^I \eta_{IJ} p^J \leq 2\kappa$.

Thus, we conclude that the theta lift \eqref{defLift} is a zero-mode of the Baston operator
\eqref{Baston}, and is given by the Penrose transform of the element in $H^1(\cZ,\cO(-2))$
defined by the formal sum of holomorphic functions (with a suitable prescription
for $p^I=0$, which we are not able to provide at this point)
\be
\label{HPsi}
\nu H_{(-1)} = \sum_{p^I} \frac{\Omega_{\rm H}(p^I)}{p^I \eta_I} \,
\exp\left( \pm 2 \pi \I\, p^I \mu_I / h \right).
\ee
For $n=4$ and $\Phi(\tau)=1$,  it is known that $\Psi$ is the
automorphic form attached to the minimal representation of $SO(4,4)$ \cite{Kazhdan:2001nx},
which can be realized in a submodule of $H^1(\cZ,\cO(-2))$ \cite{MR1421947}.
Its Fourier coefficients $\Omega_{\rm H}(p^I)$ have support on the cone $p^I \eta_{IJ} p^J=0$,
and are given by the divisor sum $\sum_{k|p^I} k$, which is simply related
to the spherical vector of the minimal representation \cite{Kazhdan:2001nx,MR2094111}.
The result \eqref{HPsi} shows that a much wider class of automorphic forms under
$SO(4,n,\IZ)$ can be realized in twistor space as sections of $H^1(\cZ,\cO(-2))$. Since their
Fourier coefficients are no longer restricted to the cone $p^I \eta_{IJ} p^J=0$, they are no
longer attached to the minimal representation, but presumably to the more general class
of quaternionic discrete series representations (or continuations thereof) \cite{MR1421947}.
(See e.g. \cite{Gunaydin:2007qq} for a pedestrian view on quaternionic discrete representations
for rank 2 reductive groups).
From a mathematical viewpoint, it would be very interesting
to construct explicit examples of $SO(4,n,\IZ)$ automorphic forms
attached to general quaternionic discrete representations.

\providecommand{\href}[2]{#2}\begingroup\raggedright\endgroup

\end{document}